\documentclass[prd,twocolumn,showpacs,preprintnumbers,floats,floatfix,nofootinbib]{revtex4-1}
\usepackage{graphicx}
\usepackage{dcolumn}
\usepackage{latexsym}

\begin{document}

\title{Three dimensional distorted black holes: using the Galerkin-Collocation method}
\author{H. P. de Oliveira}
\email{oliveira@dft.if.uerj.br}
%\author{E. L. Rodrigues}
%\email{elrodrigues@uerj.br}
\affiliation{{\it Universidade do Estado do Rio de Janeiro}\\
{\it Instituto de F\'{\i}sica - Departamento de F\'{\i}sica Te\'orica}\\
{\it Cep 20550-013, Rio de Janeiro, RJ, Brazil}}

\author{E. L. Rodrigues}
\email{eduardo.rodrigues@unirio.br}
\affiliation{{\it Universidade Federal do Estado do Rio de Janeiro}\\
{\it Centro de Ci\^encias Exatas e Tecnologia - Departamento de Inform\'atica Aplicada}\\
{\it Cep 22290-040, Rio de Janeiro, RJ, Brazil}}

\date{\today}

\begin{abstract}
We present an implementation of the Galerkin-Collocation method to determine the initial data for non-rotating distorted three dimensional black holes in the inversion and puncture schemes. The numerical method combines the key features of the Galerkin and Collocation methods which produces accurate initial data. We evaluated the ADM mass of the initial data sets, and we have provided the angular structure of the gravitational wave distribution at the initial hypersurface by evaluating the scalar $\Psi_4$ for asymptotic observers.
\keywords{Initial data; spectral methods.}
\end{abstract}

%\pacs{04.25.D-, 04.30-w}

\maketitle

%---------------------------------------------------------------------------------------------------------------------------------------------------------------------------------------------------------------------------------------
\section{Introduction}
%---------------------------------------------------------------------------------------------------------------------------------------------------------------------------------------------------------------------------------------

The full 3D evolution of the Einstein field equations figures as the most-challenging task for numerical relativity despite the progress achieved so far \cite{pfeiffer_3d}. In order to evolve any 3D code, one needs to specify the initial data representing a physically relevant system. Among all possible configurations, those involving black holes are of interest. The strong gravitational fields produce the ideal arena in which the fully general relativistic effects take place.

In this direction, there is a class of the black hole initial data known as distorted black holes that Bernstein et al \cite{bernstein} introduced. They assumed the axisymmetric initial that consists in a black hole with or without rotation in interaction with a cloud of gravitational waves of variable intensity about the black hole. Later, Brandt et al \cite{brandt} relaxed the axisymmetry and considered the most general three-dimensional distorted black holes. An important motivation in establishing and evolving distorted black holes is to reproduce the late stages of binary black hole coalescence. In addition, the dynamics of distorted black holes can provide a simple framework to study in details the efficiency of gravitational wave extraction, together with the determination of wave templates perceived by a distant observer.

Recently, we have applied the Galerkin-Collocation spectral method \cite{deol_rod_idata} to determine accurately two initial data sets for numerical relativity: pure Brill waves, and axisymmetric non-rotating distorted black holes. These problems were considered previously in the realm of traditional pseudospectral \cite{kidder} and finite difference methods \cite{id_finite_dif}. Several relevant works dealing with pseudospectral codes for the determination of single black hole initial data can be found in Refs. \cite{pfeiffer_CPC,ansorg_1,pfeiffer_thesis,pfeiffer_bh_gw,schinkel}.

There are two main strategies to describe the initial data sets for single and multiple black holes. We mention the use of isometry conditions at the inner boundaries in order to represent the black holes throats \cite{bernstein}. Another approach is  the puncture method proposed by Brandt and Brugmann \cite{brandt_brugmann}. This method proved to be very effective in describing initial data for multiple black holes, in particular binary black hole systems. Basically, it consists in splitting the conformal factor of the spatial metric into singular and non-singular terms. Brown and Lowe \cite{brown_lowe} applied the puncture method for the determination of distorted black holes spacetimes with the implementation of adaptive mesh refinement in their finite difference code to solve the elliptic equation resulting from the Hamiltonian constraint. As they have pointed out, it was necessary to perform the computation on a large grid with high resolution near the black hole. The puncture data for binary black holes or neutron stars using single and multi-domain spectral methods were considered by Grandclement et al \cite{grandclement}, Ansorg et al \cite{ansorg_1,ansorg_bbh,ansorg_07}, Pfeiffer \cite{pfeiffer_thesis,pfeiffer2}, Foucart et al \cite{foucart}, Ruchlin at al \cite{ruchlin}, Lovelace et al \cite{lovelace} and Koutarou at al \cite{koutarou}.

The main goal of the present work is to apply the Galerkin-Collocation method \cite{deol_rod_bondi,deol_rod_RT,deol_rod_idata} to obtain three-dimensional distorted black holes initial data sets. We have developed algorithms in the realm of the inversion method and the puncture method with domain decomposition. We have organized the paper as follows. Section II presents briefly the basic equations of the $3+1$ formulation for the initial data problem in both inversion and puncture methods. Section III is devoted the describe the numerical implementation of the Galerkin-Collocation method in both methods, where the choice of basis functions that satisfy the boundary conditions constitutes the cornestone of the codes. The spherical harmonics are the most natural basis functions for the angular domain, whereas the radial basis functions are expressed as suitable linear combinations of the Chebyshev polynomials. The condition of inversion symmetry had to be satisfied by imposing a relation between the unknown modes. We have implemented the puncture method with a simple version of the domain decomposition that consists in dividing the spatial domain into several regions, wherein each region we solve the Hamiltonian constraint and match these solutions across the domains. Section IV shows the convergence tests together with the asymptotic behavior of the spin-weighted scalar $\Psi_4$ which provides the pattern of the gravitational field associated to the distorted black hole at the initial slice. Finally, in Section V we  make some concluding remarks.

%---------------------------------------------------------------------------------------------------------------------------------------------------------------------------------------------------------------------------------------
\section{The initial data problem: basic equations}
%---------------------------------------------------------------------------------------------------------------------------------------------------------------------------------------------------------------------------------------

The basic equations for the initial data we are going to solve arise from the 3+1 formulation \cite{adm,york} of the Einstein's field equations. The initial data cannot be specified arbitrarily, but it must satisfy in vacuum four constraint equations given by,

\begin{eqnarray}
^{(3)}R + K^2 - K_{ij}K^{ij} &=& 0 \label{eq1}\\
\nonumber\\
^{(3)}\nabla_j\,(K^{ij} - K \gamma^{ij}) &=& 0 \label{eq2},
\end{eqnarray}

\noindent where $\gamma_{ij}$ and $K_{ij}$ are the metric and extrinsic curvature of the 3-dimensional spacelike hypersurfaces that foliate the spacetime, respectively. All quantities are evaluated on the 3-dimensional hypersurfaces, and $K = \gamma_{ij} K^{ij}$. These four equations are known as the Hamiltonian and momentum constraints, respectively.

Following Brandt et al \cite{brandt} we are going to consider the initial data at the moment of time symmetry, meaning that the extrinsic curvature is zero, $K_{ij}=0$ at the initial slice or hypersurface. In this case, three constraint equations vanish identically remaining  the Hamiltonian constraint $^{(3)}R=0$ which fixes the three-metric or initial data. It is appropriate to follow the York-Lichnerowicz \cite{york} approach expressing the metric $\gamma_{ij}$ in conformal form,

\begin{equation}
\gamma_{ij} = \Psi^4 \bar{\gamma}_{ij}, \label{eq3}
\end{equation}

\noindent where $\Psi$ is the conformal factor and the metric $\bar{\gamma}_{ij}$ are given. The Hamiltonian constraint becomes,

\begin{equation}
\bar{\nabla}^2\,\Psi - \frac{1}{8} \bar{R} \Psi = 0. \label{eq4}
\end{equation}

\noindent Here, $\bar{\nabla}^2$ and $\bar{R}$ are the Laplace operator and the Ricci scalar associated to the metric $\bar{\gamma}_{ij}$, respectively. Therefore, the Hamiltonian constraint (\ref{eq1}) becomes an elliptic equation for the conformal factor $\Psi$ whose solution determines the initial data or the initial metric $\gamma_{ij}$.

The metric of the initial hypersurface corresponding to a three-dimensional distorted black hole is expressed as \cite{bernstein,brandt},

\begin{equation}
ds^2 = \Psi^4 \big[\mathrm{e}^{2q}(dr^2+r^2d\theta^2)+r^2 \sin^2 \theta d\phi^2\big], \label{eq5}
\end{equation}

\noindent where the function $q=q(r,\theta,\phi)$ represents the distribution of gravitational wave amplitude \cite{brill} that satisfy certain boundary conditions to ensure the regularity and the asymptotic flatness of the metric. These boundary conditions are:

\begin{eqnarray}
q(r,0,\phi)=q(r,\pi/2,\phi)=0,\;\;\ \lim_{r \rightarrow \infty} q = \mathcal{O}(r^{-2}). \label{eq6}
%\left\{
%        \begin{array}{l l}
%q(0,z)=0,\;\frac{\partial q}{\partial \rho}(0,z)=0,\;\;\left(\frac{\partial q}{\partial z}\right)_{z=0}=0,\\
% \\
%q = \mathcal{O}(r^{-2}),\; \mathit{as\,\,r \rightarrow \infty} \\
%\\
%\left(\frac{\partial \Psi}{\partial \rho}\right)_{\rho=0}=0,\;\;\left(\frac{\partial \Psi}{\partial z}\right)_{z=0}=0 \\
%\\
%\Psi =1 +\frac{M}{2r} + \mathcal{O}(r^{-2}), \; \mathit{as\,\,r \rightarrow \infty},
%\end{array} \label{eq6}
%\right.
\end{eqnarray}

\noindent We have considered the gravitational wave amplitude distribution function introduced by to Bernstein et al \cite{bernstein},

\begin{eqnarray}
%q = A_0 \sin^n\theta\,f(r)(1+c\cos^2 \phi). \label{eq7}
q(r,\theta,\phi) = A_0 \sin^n\theta\left[\mathrm{e}^{-\left(\frac{\eta+\eta_0}{\sigma}\right)^2}+ \mathrm{e}^{-\left(\frac{\eta-\eta_0}{\sigma}\right)^2}\right](1+c\cos^2 \phi),\nonumber \\
 \label{eq7}
\end{eqnarray}

\noindent where $A_0$ denotes the amplitude of the Brill wave \cite{brill}, the free parameter $c$ indicates the deviation from axisymmetry, $\eta = \ln(r/a)$ and $n \geq 2$ is an even integer; $\eta_0,\sigma$ are constants associated to the position and width, respectively, of the Brill wave. %The second expression for $f(r)$ is a variation of the Eppley \cite{eppley} data for which $f(r)=(1+(r/\lambda)^{s})^{-1}$ and $s \geq 2$.

The conformal factor must satisfy the condition,

\begin{equation}
\Psi = 1 +\frac{\mathcal{M}}{2r} + \mathcal{O}(r^{-2}) \label{eq8}
\end{equation}

\noindent at $r \rightarrow \infty$ as a consequence of the Robin boundary condition asymptotically, %guarantying the asymptotically flatness of the spacetime.
and parameter $\mathcal{M}$ is the ADM mass. The inner boundary is placed at the throat of the black hole and satisfies the isometry condition condition,

\begin{equation}
\left(\frac{\partial \Psi}{\partial r}+\frac{\Psi}{2a}\right)_{r=a}=0, \label{eq9}
\end{equation}

\noindent where $a=M_0/2$, and $M_0$ is the mass of the black hole that results from setting $q=0$. Therefore, the Bernstein data sets are obtained after solving the Hamiltonian constraint in the region $r \geq a$ satisfying the boundary conditions (\ref{eq6}) and (\ref{eq8}).

An alternative way of constructing isometric distorted black hole data sets is provided by the so-called puncture method \cite{brandt_brugmann}. The central idea of the puncture method is to split the conformal factor into a singular and nonsingular terms, and to consider the whole spatial domain instead of being restricted to the region outside the throat of the hole. Accordingly, the conformal factor is written as,

\begin{equation}
\Psi = u + \frac{m}{2 r}, \label{eq10}
\end{equation}

\noindent where $m$ is a new parameter of the method. Notice that the term $m/2r$ is singular at the origin, whereas the function $u$ is the nonsingular term.

We present the Hamiltonian constraint expressed in function of $u$ after substituting the decomposition (\ref{eq10}) into the Eq. (\ref{eq4}), which results in,

\begin{equation}
\bar{\nabla}^2\,u - \frac{1}{8} \bar{R} u = \frac{m}{16} \bar{R}.  \label{eq11}
\end{equation}

\noindent This equation is identical to Eq. (\ref{eq4}) with an addition source term $16 m/\bar{R}$. Brown and Lowe \cite{brown_lowe} have shown that the data sets obtained after solving Eq. (\ref{eq11}) satisfy automatically the isometry condition if $m = M_0 = 2 a$. Furthermore, it is possible to generate initial spacetimes that do not satisfy the isometry condition by setting $m \neq M_0$.

%---------------------------------------------------------------------------------------------------------------------------------------------------------------------------------------------------------------------------------------
\section{The numerical scheme}
%---------------------------------------------------------------------------------------------------------------------------------------------------------------------------------------------------------------------------------------

%--------------------------------
\subsection{The inversion method}
%--------------------------------

We follow the procedure we have employed in Ref. \cite{deol_rod_idata} to deal with the axisymmetric distorted black hole data sets. The starting point is to establish an approximate expression for the conformal factor given by,

\begin{equation}
\Psi _{a}(r,\theta,\phi) = 1 + \sum^{N_x,N_y}_{k,l = 0}\sum^{l}_{m=-l}\,c_{klm} ~\chi_{k}(r) Y_{lm}(\theta,\phi), \label{eq12}
\end{equation}

\noindent where $c_{klm}$  represents the unknown modes and $N_x,N_y$ are the truncation orders that limit the number of terms in the above expansion. The angular patch has the spherical harmonics, $Y_{lm}(\theta,\phi)$, as the basis functions, whereas the radial basis functions, $\chi_k(r)$, are the same used for the axisymmetric case,

\begin{eqnarray}
\chi_k(r) = \frac{1}{2}(TL_{k+1}(r)-TL_k(r)),\label{eq13}
\end{eqnarray}

\noindent where, the rational Chebyshev polynomials $TL_k(r)$ are defined according to \cite{boyd},

\begin{eqnarray}
TL_k(r) = T_k\left(x=\frac{r-a-L_r}{r-a+L_r}\right),\label{eq14}
\end{eqnarray}

\noindent where $T_k(x)$ is the traditional Chebyshev polynomials. The radial domain $a \leq r < \infty$ is equivalent to $-1 \leq x \leq 1$ (see Fig. 1) with $L_r$ being the map parameter whose convenient choice improves the accuracy of the approximate solution. It can be shown that $\chi_k(r) = \mathcal{O}(r^{-1})$ as $r \rightarrow \infty$ reproduces the boundary condition $\Psi(r,\theta) = 1 + \mathcal{O}(r^{-1})$.

\begin{figure*}[htb]
\includegraphics[scale=0.27]{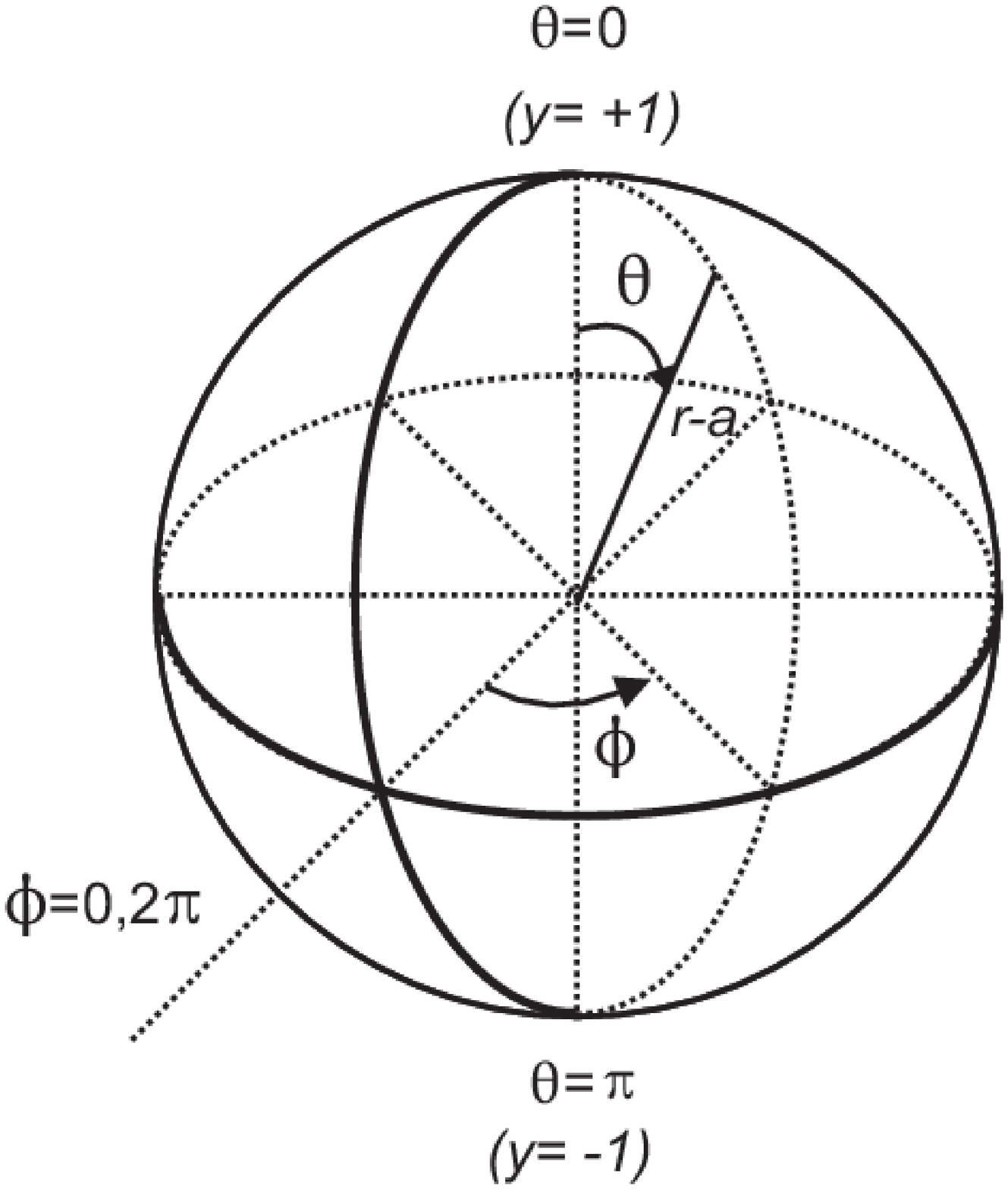} \includegraphics[height=5.6cm,width=5.3cm]{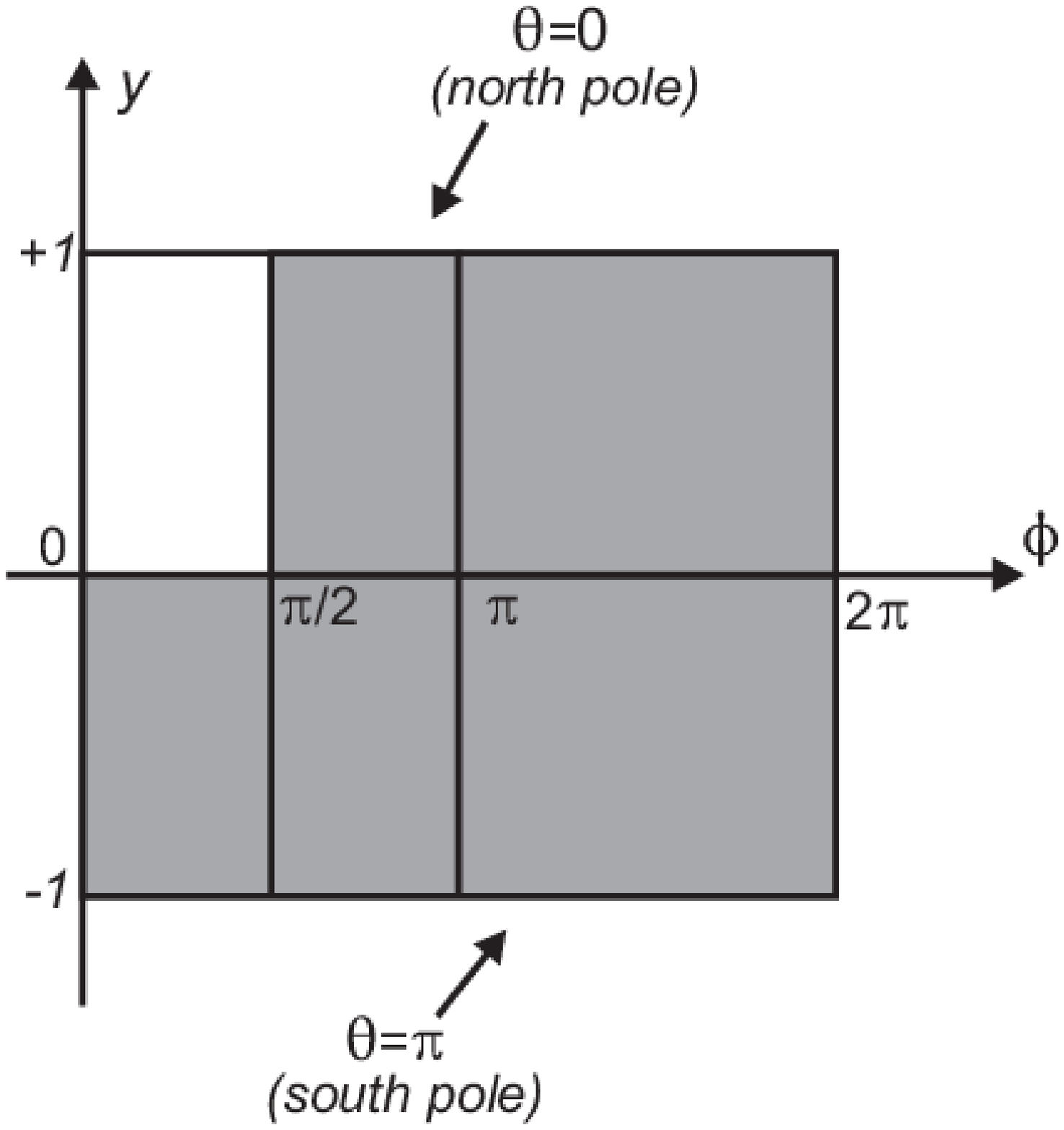}\includegraphics[height=5.3cm,width=5.3cm]{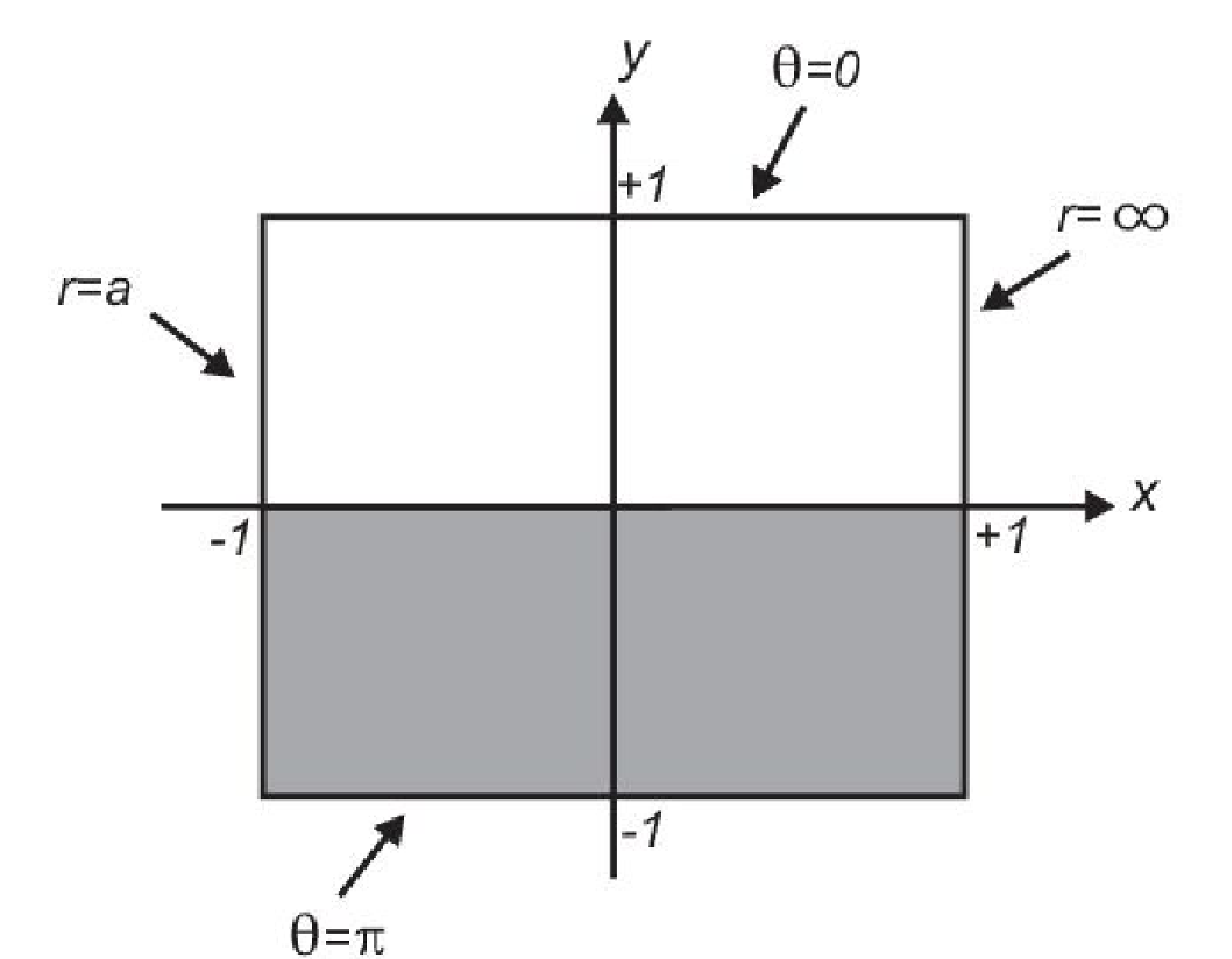}%[scale=0.3]{grid2}%\includegraphics[scale=0.35]{fig1c}
\caption{The angular variables $(\theta,\phi)$ cover the surface of the sphere, or the subdomain $r=\mathrm{constant}$ (figure on the top). We have covered the entire spatial domain with the coordinates  $(x,y,\phi)$. We have shown in the left and right panels the subdomains $x=\mathrm{constant}$ ($r=\mathrm{constant}$) and $\phi=\mathrm{constant}$, respectively. Notice that the spatial infinity ($r \rightarrow \infty$) is placed at $x = 1$. The initial data has equatorial plane symmetry due to the form of the gravitational wave amplitude with $n$ even (see Eq. (\ref{eq9}))  ($0 \leq \theta \leq \pi/2$ or $-1 \leq y \leq 1$). In addition, the angular dependence on $\phi$ allows us to consider only the patch $0 \leq \phi \leq \phi/2$.}
%\vspace{-2cm}
\end{figure*}

The spherical harmonics are complex functions implying that the modes $c_{klm}$ must be complex, but they satisfy certain symmetry relations in order to produce a real conformal factor. These symmetry relations are,

\begin{equation}
c^*_{kl-m} = (-1)^{-m} c_{klm}, \label{eq15}
\end{equation}

\noindent since $Y^*_{l-m} = (-1)^{-m} Y_{lm}$. As a consequence, the total number of independent unknown coefficients is $(N_x+1) \times (N_y+1)^2$.

The inversion symmetry condition (\ref{eq9}) is satisfied in an approximate way according to,

%\small{
\begin{eqnarray}
\left<\left(\frac{\partial \Psi}{\partial r}+\frac{\Psi}{2a}\right)_{r=a},Y_{lm}\right> &=& \int_{\Omega}\,\left(\frac{\partial \Psi}{\partial r}+\frac{\Psi}{2a}\right)_{r
=a} Y^*_{lm}(\theta,\phi) d \Omega \nonumber \\
&=& 0, \label{eq16}
\end{eqnarray}
%}

\noindent for all $l=0,1,..,N,\;m=-l,..,l$. The integrals are evaluated using quadrature formulas, which is typical of the G-NI (Galerkin with numerical integration) method \cite{canuto}:

{\small
\begin{eqnarray}
\int_{\Omega}\,\left(\frac{\partial \Psi}{\partial r}+\frac{\Psi}{2a}\right)_{a} Y^*_{lm}(\theta,\phi) d \Omega &\simeq& \sum_{j,k=0}^{N_1,N_2} \left(\frac{\partial \Psi}{\partial r}+\frac{\Psi}{2a}\right)_{(a,\theta_j,\phi_k)} \times \nonumber\\
& &  Y^*_{lm}(\theta_j,\phi_k) w_j v_k = 0, \label{eq17}
\end{eqnarray}
}

\noindent where $(\theta_j,\phi_k)$ with $j=0,1,..,N_1, k=0,1,..,N_2$, respectively, are the collocation points on the angular domain and given by,

\begin{eqnarray}
\theta_j&=&\arccos (y_j),\; \mathrm{and}\;\;y_j=-1,\,\mathrm{zeros\,of}\, \frac{d P_{2 N_1}}{d y}, 1 \label{eq18}\\
\phi_k &=& \frac{2 \pi k}{N_2+1}. \label{eq19}
\end{eqnarray}

\noindent The quantities $w_j, v_k$ are the corresponding weights \cite{fornberg}, and we have chosen $N_{1}=N_{2}=2 N_y$ for better accuracy in the calculation of the integrals (see also Ref. \cite{deol_rod_RT,peyret}). We have obtained $(N_y+1)^2$ linear algebraic relations for the unknown coefficients $c_{klm}$ from (\ref{eq17}) that can be solved to express the $(N_y+1)^2$ coefficients $c_{N_x l m}$ in function of the remaining ones. Therefore, we ended up with a total of $N_x \times (N_y+1)^2$ independents coefficients. The use of the new coordinates $x,y$ together with $\phi$ covers the entire spatial domain as illustrated in Fig. 1.

The residual equation associated to the Hamiltonian constraint is obtained by substituting the approximate conformal factor $\Psi(r,\theta,\phi)$, into Eq. (\ref{eq4}). We represent this equation by $\mathrm{Res}(r,\theta,\phi)$ recognizing that it does not vanish identically due to the approximated conformal factor. Next, the coefficients are determined such as to force the residual equation to zero in an approximate sense \cite{finlayson} as we have done with the inversion symmetry condition. Then, it follows that,

\begin{eqnarray}
\left<\mathrm{Res},\xi_k(r) Y_{lm}\right> =&& \int_{a}^\infty\,\xi_k(r) dr\,\times \\ \nonumber
&& \underbrace{\int_{\Omega}\,\mathrm{Res}(r,\theta,\phi) Y^*_{lm}(\theta,\phi) d \Omega}_{\mathrm{Res}_{lm}(r)} = 0, \label{eq20}
\end{eqnarray}

\noindent where the $\xi_k(r)$ are known as the test functions \cite{finlayson}, and $l=0,1,..,N_y,\,m=-l,-l+1,..,l$ and $k=1,2,..,N_x$. As before we have evaluated the integrals over the angular domain  using the Gauss quadrature formulas, or,

\begin{equation}
\mathrm{Res}_{lm}(r) \approx \sum_{j,k=0}^{N_1,N_2} \mathrm{Res}(r,\theta_j,\phi_k) Y^*_{lm}(\theta_j,\phi_k) w_j v_k. \label{eq21}
\end{equation}

\noindent Now, by choosing test functions as delta of Dirac functions $\xi_k(r)=\delta(r-r_k)$, where $r_k$ represents the collocation points on the radial patch, we obtain the set of equations for the independent coefficients $c_{klm}$ expressed as,

\begin{equation}
\left<\mathrm{Res},\xi_k(r) Y_{lm}\right> = \mathrm{Res}_{lm}(r_k)=0, \label{eq22}
\end{equation}

\noindent where $k=1,2,..,N_x$. The radial collocation points $r_k$ are,

\begin{eqnarray}
r_k &=& a + L_r \frac{(1+x_k)}{1-x_k},\; \mathrm{where} \nonumber \\
x_k&=&\cos\left(\frac{k \pi}{N_x}\right). \label{eq23}
\end{eqnarray}

%\noindent The set of equations resulting from (\ref{eq22}) has $N_x\times(N_y+1)^2$ linear algebraic equations for the same number of unknown coefficients $c_{klm}$. The number of independent coefficients was further reduced taking into account the symmetries of the gravitational wave amplitude (cf. Eq. (\ref{eq7})). The resulting equations are solved using standard linear solvers of Maple or Matlab, determining the coefficients and consequently the approximate conformal factor.

\noindent The set of equations (\ref{eq22}) has $N_x\times(N_y+1)^2$ linear and ill-conditioned algebraic equations for the same number of unknown coefficients $c_{klm}$. The preconditioning technique (see Ref. \cite{pfeiffer_CPC} and references therein) allows to reduce the number of iterations in solving ill-conditioned linear systems especially those with an enormous number of equations. In the present case we have reduced further the number of independent coefficients by taking into account the symmetries of the gravitational wave amplitude (cf. Eq. (\ref{eq7})). Therefore, the resulting equations are solved using standard linear solvers of Maple or Matlab, determining the coefficients and consequently the approximate conformal factor.

%The preconditioning technique allows to reduce the number of iterations in solving ill-conditioned linear systems especially those with an enormous number of equations. In our case, we have included the additional symmetries due to the choice of the function $q$ and the number of equations (or modes) were reduced significantly. For instance, by taking $N_x=70$ and $N_y=6$ (for the inversion method), we have 700 equations to be solved. Then, we have applied satisfactorily the standard linear solvers of Maple or Matlab to obtain the coefficients $c_{kjl}$.

%--------------------------------
\subsection{The puncture method}
%--------------------------------

\begin{figure}[htb]
\includegraphics[scale=0.15]{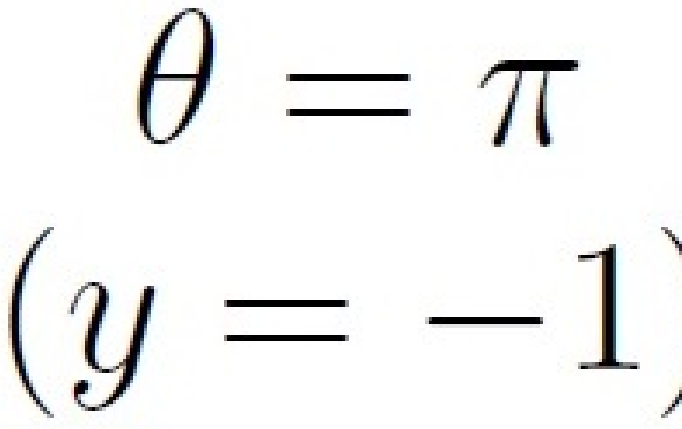}
\caption{Illustration of the regions $\mathcal{D}_1$ and $\mathcal{D}_2$ viewed from the plane $(r,\theta)$, where the bold line represents the boundary $r=r_0$.}
\end{figure}

For the Galerkin-Collocation implementation of puncture method it is likewise necessary to establish an approximate expression for the function $u(r,\theta,\phi)$. The fulfillment of the Robin condition implies that asymptotically $u = 1 + \mathcal{O}(r^{-1})$, and due to the nonsingular nature of $u$, it follows that,

\begin{equation}
u _{a}(r,\theta,\phi) = 1 + \sum^{N_x,N_y}_{k,l = 0}\sum^{l}_{m=-l}\,c_{klm} ~\chi_{k}(r) Y_{lm}(\theta,\phi). \label{eq24}
\end{equation}

\noindent This expression is identical to the approximate conformal factor given by Eq. (\ref{eq12}), where again the $c_{klm}$ represents the unknown modes, $N_x,N_y$ are the truncation orders, and the spherical harmonics are the angular basis functions. The radial basis functions are given by Eq. (\ref{eq12}), but the rational Chebysehv polynomials are given by,

\begin{eqnarray}
TL_k(r) = T_k\left(x=\frac{r-L_r}{r+L_r}\right), \label{eq25}
\end{eqnarray}

\noindent in order to cover the whole radial domain $0 < r < \infty$ being equivalent to $-1 < x <1$.

The determination of the coefficients $c_{klm}$ follows the same steps we have devised previously. Since we are interested on the isometric data sets, we have set $m=2a$ \cite{brown_lowe}.

The domain decomposition technique \cite{pfeiffer_thesis,boyd} can be implemented more naturally in the scheme of the puncture method. It consists in dividing the spatial domain into two or more distinct regions, each one with approximate expressions for the function $u$. We present here a simple, but efficient version of the domain decomposition by establishing two regions: the first region $\mathcal{D}_1: 0 < r \leq r_0$ and the second region defined by $\mathcal{D}_2: r \geq r_0$. In Fig. 2 we present the general scheme of the domain decomposition viewed in the plane $(r,\theta)$, where the bold line corresponds to the boundary $r=r_0$ separating the regions $\mathcal{D}_1$ and $\mathcal{D}_2$. Naturally, the junction conditions,

\begin{eqnarray}
u^{(1)}(r_0,\theta,\phi)=u^{(2)}(r_0,\theta,\phi),\;\frac{\partial u^{(1)}}{\partial r}\Big{|}_{r_0}=\frac{\partial u^{(1)}}{\partial r}\Big{|}_{r_0},\label{eq26}
\end{eqnarray}

\noindent must be satified at the boundary $r=r_0$. Numerically these relations are approximated, in the same way, as described by Eqs. (\ref{eq16}) and (\ref{eq17}). In Table 1, we summarize the approximate functions $u$ at these regions in which the radial basis functions are chosen conveniently in each region, whereas the angular basis functions remain the same. These expansions have the spherical harmonics as the angular basis functions with the same truncation order $N_y$, but the truncation orders of the radial sector may be different in each region.

\begin{table*}
	\centering
		\begin{tabular}{c|c}
		\hline
		$D_1: 0 < r \leq r_0$ & $D_2: r \geq r_0$\\
		\hline
		\hline
		\\
		$u^{(1)}_{a}(r,\theta,\phi) = 1 + \sum\,c^{(1)}_{klm} TL_{k}(r) Y_{lm}(\theta,\phi)$ & $u^{(2)}_{a}(r,\theta,\phi) = 1 + \sum\,c^{(2)}_{klm} \chi_{k}(r) Y_{lm}(\theta,\phi)$\\
		\\
		Radial basis function:& Radial basis function:\\
	\\
		$TL_k(r) = T_k\left(x=\frac{2r}{r_0}-1\right)$ & $\chi_k(r) = \frac{1}{2}(TL_{k+1}(r)-TL_k(r))$\\
		\\
		                                          & $TL_k(r) = T_k\left(x=\frac{r-r_0-L_r}{r-r_0+L_r}\right)$\\
	\hline
																						
		\end{tabular}
		\caption{Approximate functions $u(r,\theta,\phi)$ defined at the regions $\mathcal{D}_1$ and $\mathcal{D}_2$. Notice that the radial basis functions are distinct in each region according to the boundary conditions. In the first region, we have considered the redefined Chebyshev polynomial functions with the map $x=2r/r_0-1$ that connects $0 < r \leq r_0$ to $-1 < x \leq 1$. In the second region, we have used the same radial basis function defined for the inversion method. Although not indicated explicitly, we have adopted $N_x^{(1)}$ and $N_x^{(2)}$ as the radial truncation orders for the radial expansion in the first and second regions, respectively.}
\end{table*}

%---------------------------------------------------------------------------------------------------------------------------------------------------------------------------------------------------------------------------------------
\section{Numerical results: convergence tests, ADM mass and the angular pattern of gravitational radiation}
%---------------------------------------------------------------------------------------------------------------------------------------------------------------------------------------------------------------------------------------

We present compelling numerical experiments showing the exponential convergence of the Galerkin-Collocation implementation for solving the initial data problem of three-dimensional distorted black holes using inversion and puncture methods. We have selected three tests in this direction: the convergence of the ADM mass, the $L_2$-norm associated to the difference of the solutions corresponding to successive truncation orders, and the $L_2$-norm associated to the residual Hamiltonian constraint equation considering increasing radial resolution.

We start with the calculation of the ADM masses of distorted black holes which are evaluated more efficiently using a formula derived by \'O Murchada and York \cite{ADM_mass},

\begin{equation}
E-\bar{E}=-\frac{1}{2 \pi}\oint_\infty \nabla_\alpha \Psi dS^\alpha, \label{eq27}
\end{equation}

\noindent where $E$ is the total energy of the hypersurface while $\bar{E}$ is the energy associated to the conformal metric. As pointed out by Bernstein et al \cite{bernstein} this last term vanishes since the conformal factor decays more rapidly than $1/r$, therefore, the ADM mass is given by the integral on the rhs of the above equation. By inserting $\nabla_\alpha \Psi = (\partial \Psi/\partial r,1/r \partial \Psi/\partial \theta,1/(r \sin \theta) \partial \Psi/\partial \phi )$ the final expression for the ADM mass becomes,

\begin{eqnarray}
M_{ADM} = -\lim_{r \rightarrow \infty}\,\frac{1}{2 \pi}\,\int_0^{2 \pi}\,\int_{-1}^1\,\left(\frac{\partial \Psi}{\partial r} r^2\right) d y d\phi. \label{eq28}
\end{eqnarray}

\noindent We have evaluated the above limit without approximating the infinity to some finite radius $r=r_{\mathrm{max}}$. This feature is a consequence of defining the conformal factor in the whole spatial domain. Therefore, after obtaining the approximate conformal factor the ADM mass could be calculated by direct integration.

We have obtained the convergence of the ADM mass by calculating the difference of the ADM masses corresponding to approximate solutions with distinct truncation orders. To be more specific, we  have fixed $N_y = 6$ and established that $\delta M(N_x) = M_{ADM}(N_x+5)-M_{ADM}(N_x)$. In Fig. 3 we present the exponential decay of $\delta M$ for the inversion method and the puncture method with domain decomposition in which we have fixed $N^{(1)}_x=30$ in the first domain. In both cases the saturation occurs at approximately $N_x=70$ (cf. Fig. 3), and $A_0=\sigma=\eta_0=c=1,n=4$, but the puncture method with domain decomposition presents a better convergence rate. In the case of the puncture method, only the conformal factor defined in the second region, $r \geq a$, is used to calculate the ADM mass. The values of the ADM masses evaluated in both methods is unaffected to almost all significant digits. As a last comment, we have found that the puncture method without domain decomposition in the realm of the Galerkin-Collocation implementation is not efficient in the sense of producing a poor convergence rate.

We have noticed that in this first round of numerical experiments, the rate of convergence of the ADM mass is sensitive to the choice of the map parameter $L_r$. In both inversion and puncture method, we have found that $L_r=9.0$ is the best value and adopted hereafter. The main criterion for choosing the map parameter is to coincide it approximately with the scale of the problem under consideration (cf. Ref. \cite{boyd}, pg. 369), but some trial-and-error was inevitable.

\begin{figure}
\includegraphics[height=5.5cm,width=6.0cm]{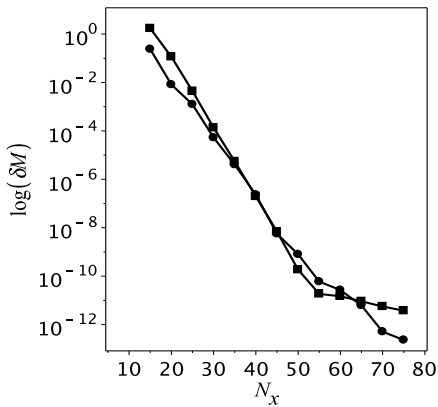}
\caption{Exponential decay of $\delta M(N_x)=M_{ADM}(N_x+5)-M_{ADM}(N_x)$ for the inversion method (boxes) and puncture method with domain decomposition (circles) with $N^{(1)}_x=30$. In both cases $N_y=6$ is fixed, as well $A_0=1,\sigma=\eta_0=c=1$. We have set $L_r=9.0$ in both methods.}
\end{figure}

For the convergence of the $L_2$-norm of $\delta \Psi$, $L_2(\delta \Psi)$, we have considered the approximate solutions as described above. The calculation of $L_2(\delta \Psi)$ for both inversion and puncture method takes into account the region outside the throat $r \geq a$. This quantity is given by,

\begin{equation}
L_2(\delta \Psi) = \left[\frac{1}{4 \pi}\,\int_0^{2 \pi} d \phi\,\int_{-1}^1 dx\,\int_0^1\,\delta \Psi(x,y,\phi)^2\,d y\right]^{1/2},\label{eq29}
\end{equation}

\noindent where $\delta \Psi(x,y,\phi)$ is obtained from (\ref{eq8}) after changing the variables $(r,\theta)$ to $(x,y)$ according to $\theta=\arccos y$ and $r=a+L_r (1+x)/(1-x)$. We have used the property $\Psi(x,y,\phi)=\Psi(x,-y,\phi)$ as a consequence of the reflection symmetry about the plane $\theta=\pi/2$ or $y=0$. The $L_2$ norm was calculated using quadrature formulae \cite{deol_rod_idata}, and its exponential decay in both methods is presented in Fig. 4. In the case of the puncture method, we have exhibited the results for distinct, but fixed values of the radial truncation order at the first region, namely $N^{(1)}_x=20,30$. In the later case the convergence is better.

For the sake of completeness, we have exhibited in In Fig. 5 the behavior of the ADM mass in function of the amplitude $A_0$ for $n=4$, $\eta_0=1$, $a=1$ and $c=1,-1,-2$. We noticed the same counterintuitive behavior of the ADM mass found in Refs. \cite{bernstein,brandt}, that is, $M_{ADM}$ initially decreases when $A_0$ increases for $A_0 \geq 0$.

\begin{figure}
\includegraphics[height=5.5cm,width=6.0cm]{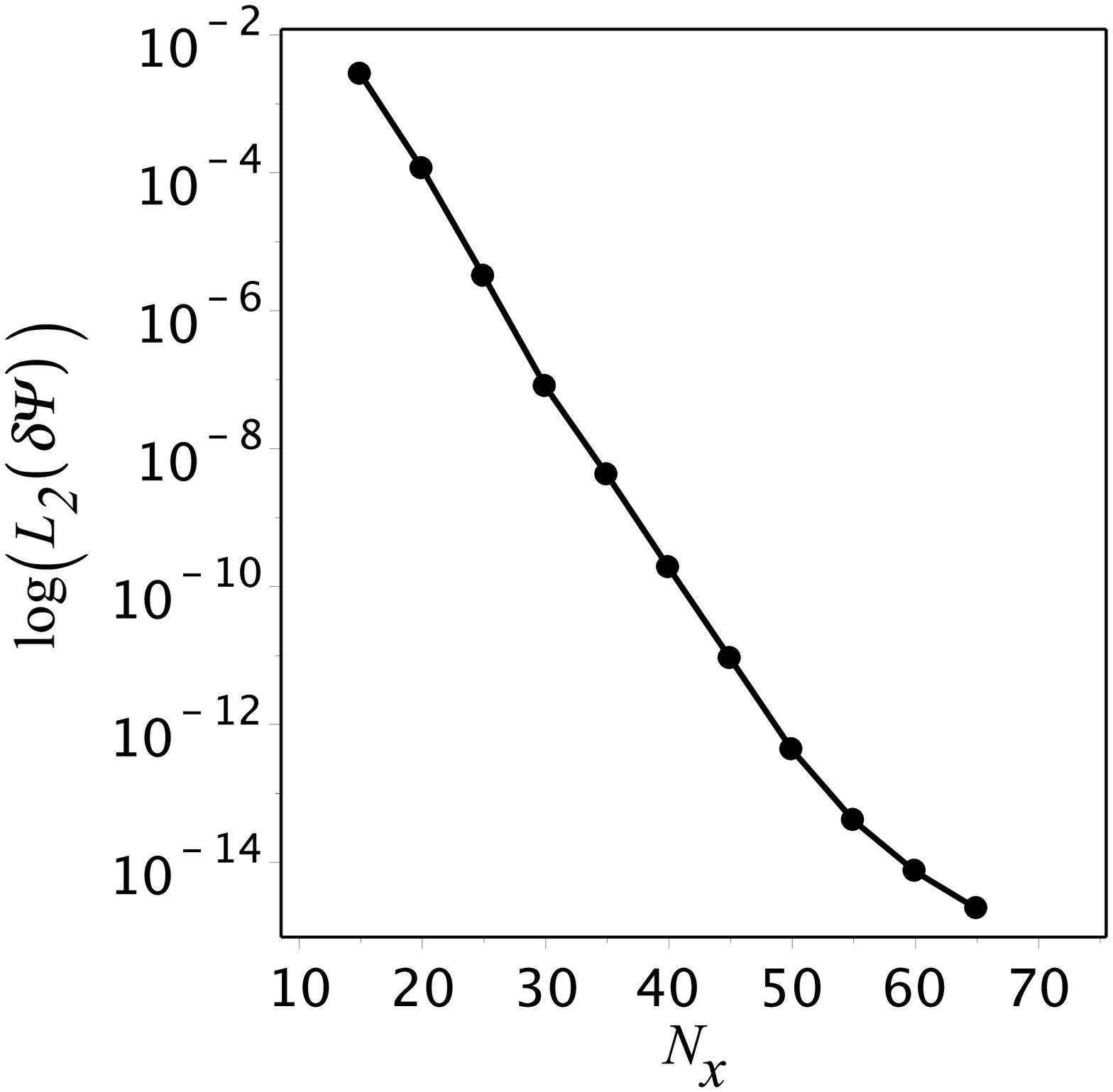}
\includegraphics[height=5.5cm,width=6.0cm]{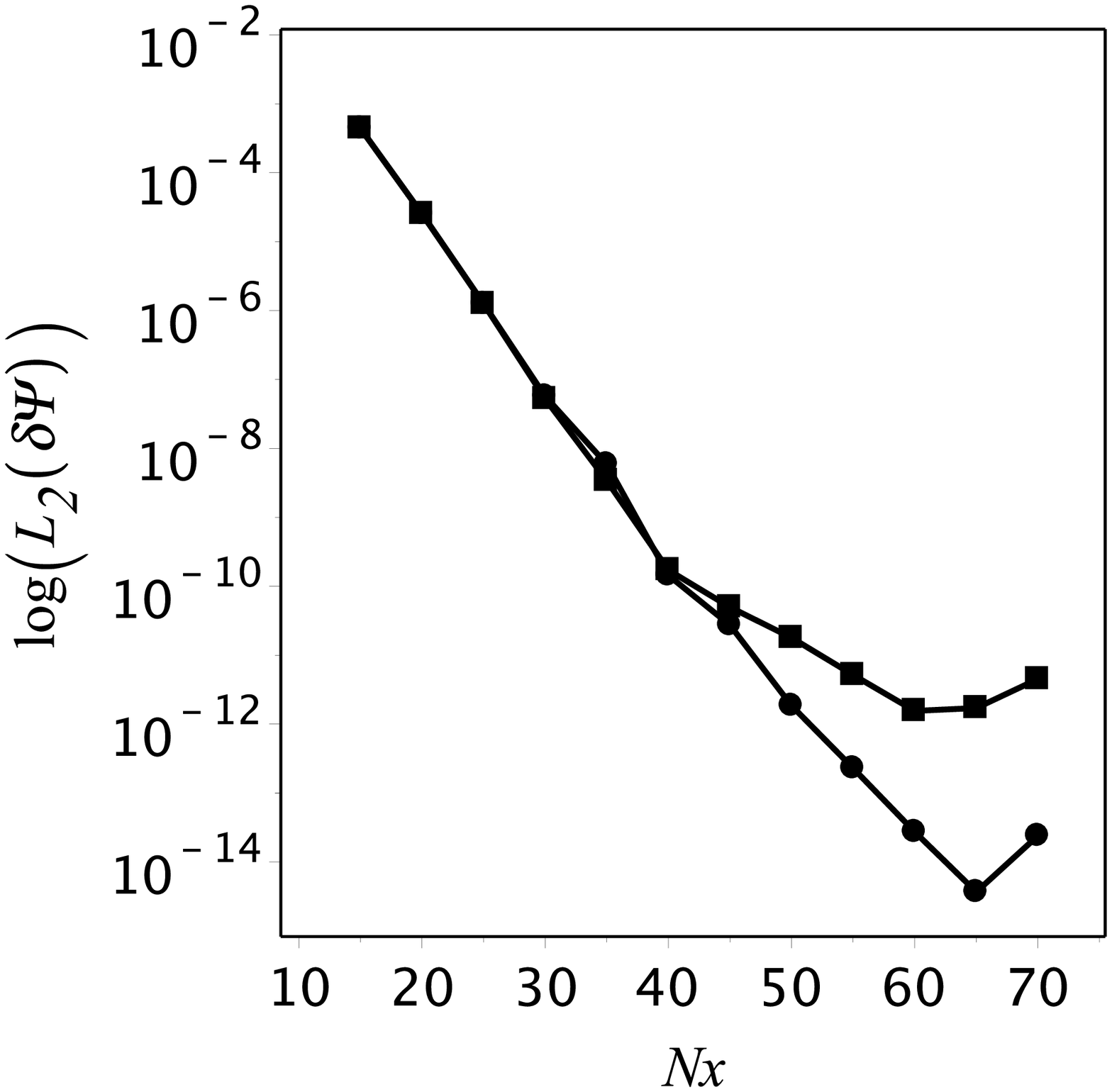}
\caption{First: $L_2(\delta \Psi)$, inversion method. We have fixed $N_y=6$ and the difference corresponds to the approximate solutions with $N_x+5$ and $N_x$; the error $L_2$ is evaluated for $r \geq a$. Second: $L_2(\delta \Psi)$, $N_x^{(1)}=20$ (box), $N_x^{(1)}=30$ (circle); puncture method with domain decomposition; $N_y=6$ remains fixed and the difference corresponds to the approximate solutions with $N_x+5$ and $N_x$; the error $L_2$ is evaluated for $r \geq a$, i.e., the second domain. In all cases $L_0=9.0$.}
\end{figure}

\begin{figure}[htb]
\includegraphics[height=4.5cm,width=5.0cm]{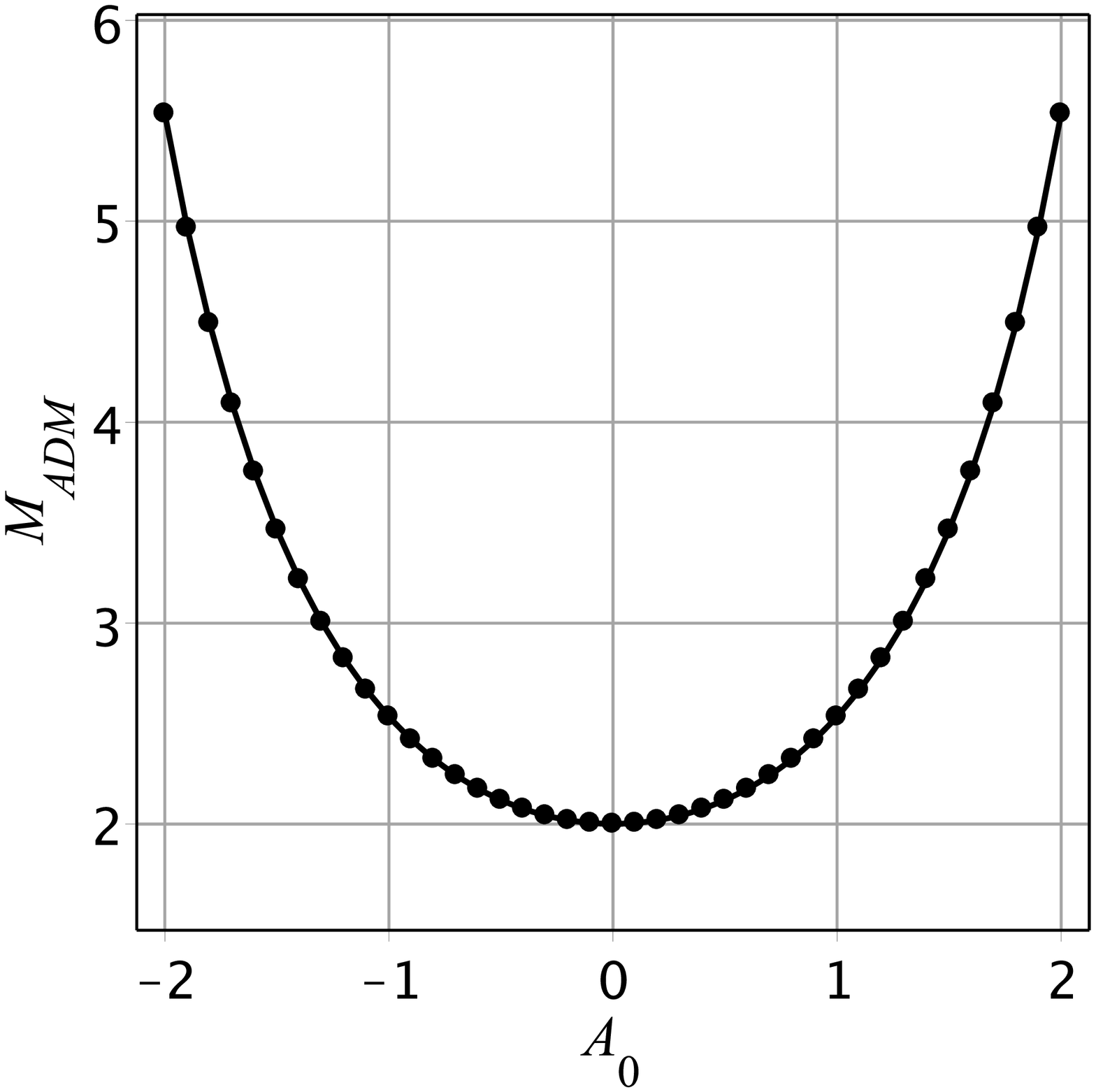}
\includegraphics[height=4.5cm,width=5.0cm]{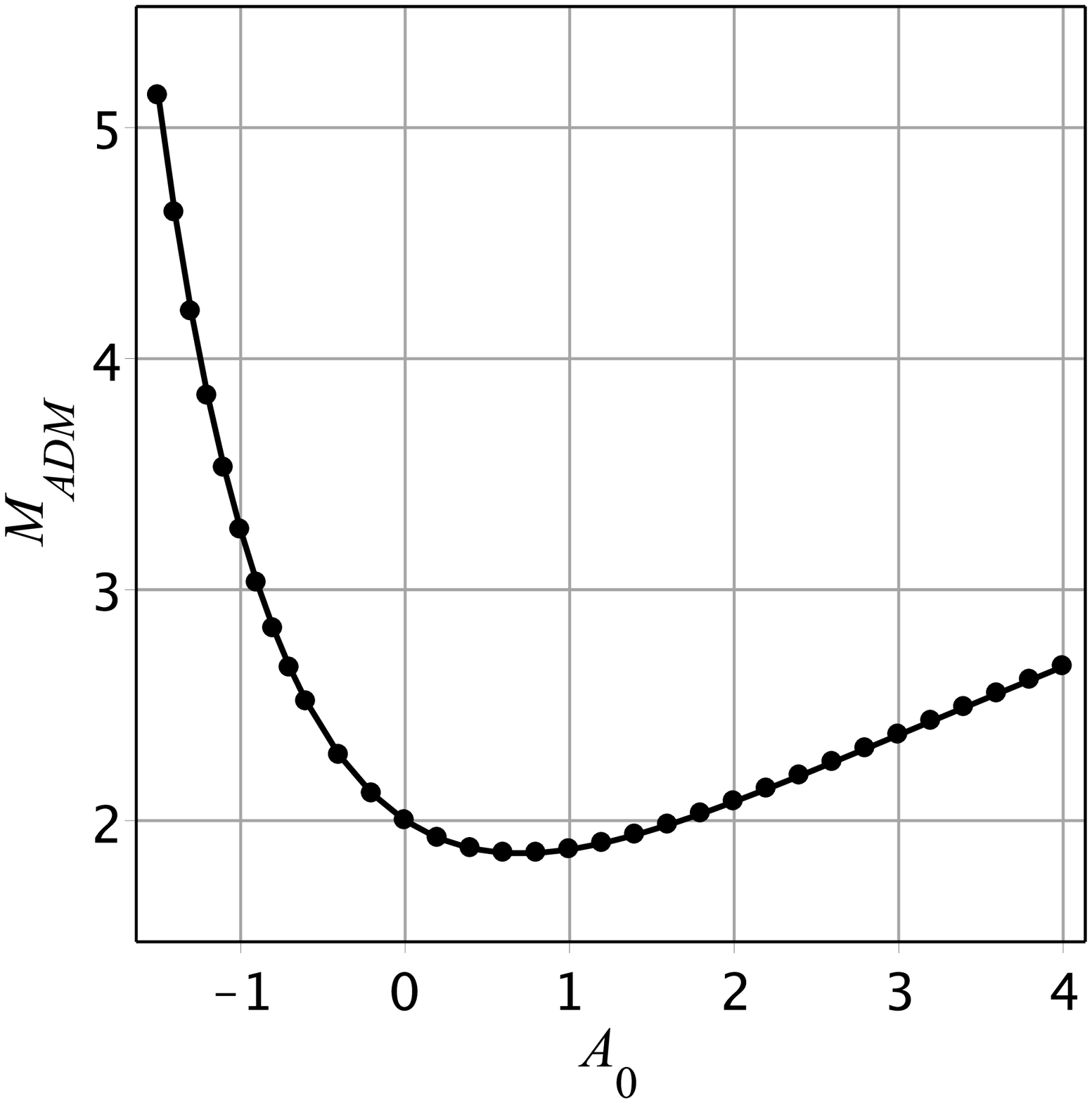}
\includegraphics[height=4.5cm,width=5.0cm]{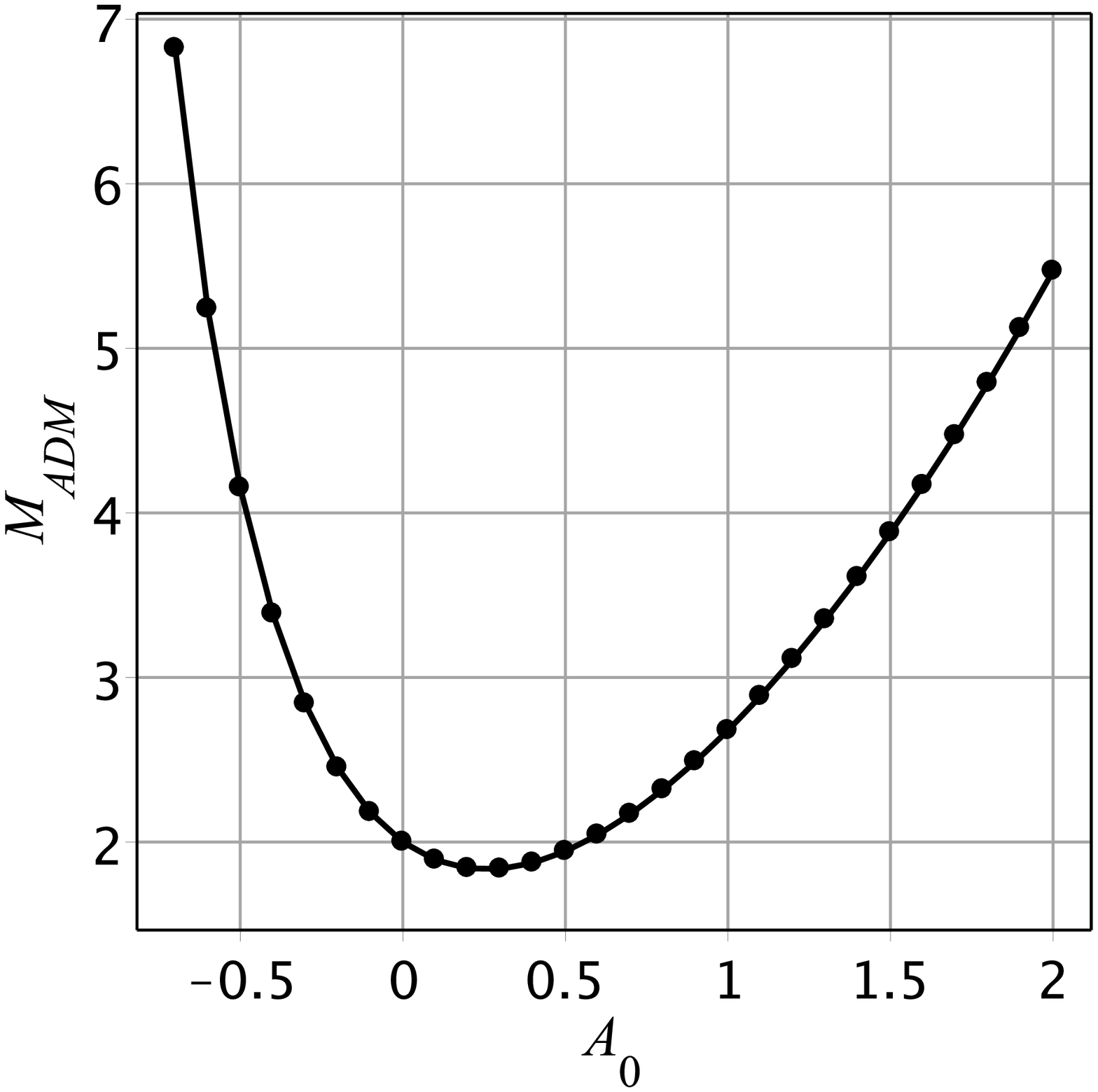}
\caption{ADM masses evaluated for the first initial data with $c=-2,-1,1$, respectively, from top to bottom.}
\end{figure}

The last numerical test is to show the convergence of the $L_2$-norm of $\mathrm{Res}(x,y,\phi)$, i. e., the residual equation associated to the Hamiltonian constraint (\ref{eq4}). We have considered the inversion method and introduced the variables $x,y$ as before. We show the results in Fig. 6 corresponding to two cases. In the first, we have fixed $N_y=10$ and increase $N_x$ for some values of the amplitude $A_0$, namely $A_0=0.01,0.5, 1.0$. Notice that $L_2(\mathrm{Res})$ achieves a limit minimum value after some value of $N_x$ that depends on $A_0$. In the second plot we have explored this aspect by fixing $A_0=1$, and for each value of $N_y$, evaluate the norm choosing $N_x$ such as to get the minimum value, according to the upper graph. For instance, as we can see from the upper graph, for $N_y=10$, we set $N_x =13$ (for $A_0 =1$). In both cases we have set $\sigma=\eta_0=c=1,n=4$, and the decay of the $L_2(\mathrm{Res})$ is indeed exponential.

\begin{figure}
\includegraphics[height=5.7cm,width=6.0cm]{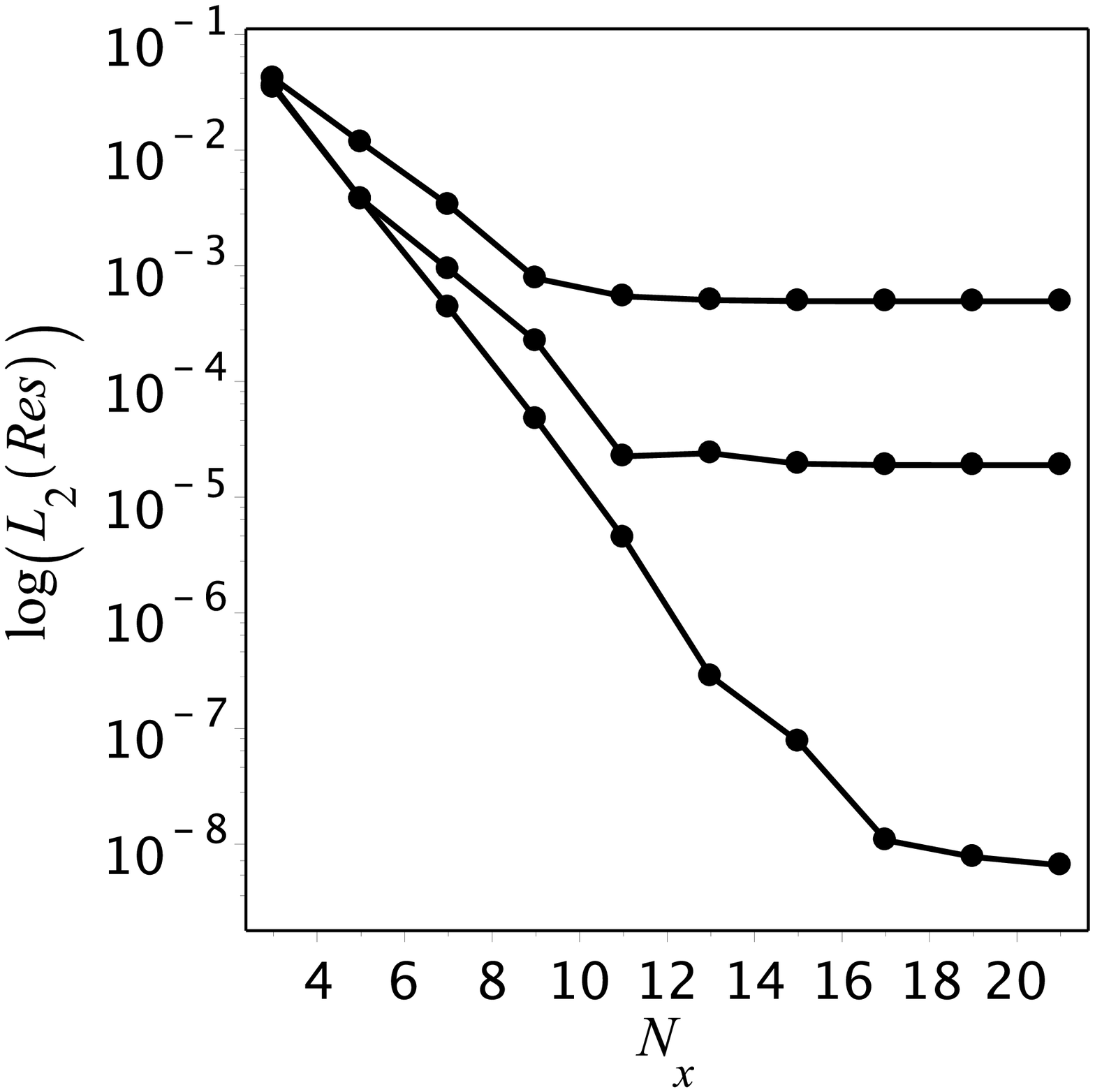}\\%[scale=0.3]
\includegraphics[height=5.7cm,width=6.0cm]{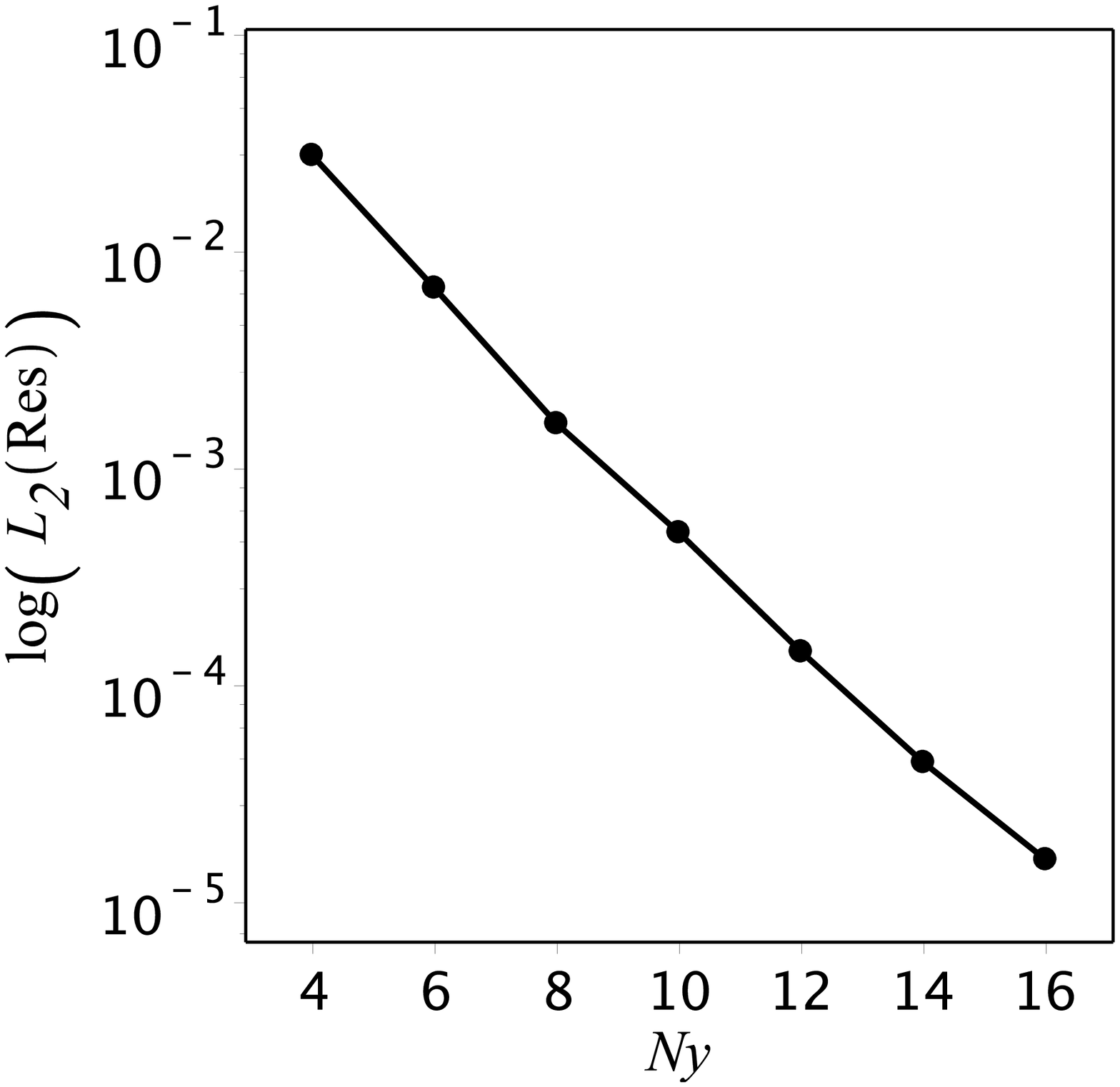}%{fig2}
\caption{Exponential decay of the $L_2$-norm associated to the residual equation in two situations. In the first, we have fixed $N_y=10$ and increased $N_x$ for three values of the amplitude, $A_0=1.0,0.5,0.01$, that corresponds to the curves from up to down. The exponential decay of the $L_2$-norm is also observed for fixed $A_0=1$ by increasing $N_y$ and $N_x$ until achieving the saturation value for the combination $N_x,Ny$.}
\end{figure}

\begin{figure}[htb]
\includegraphics[scale=0.32]{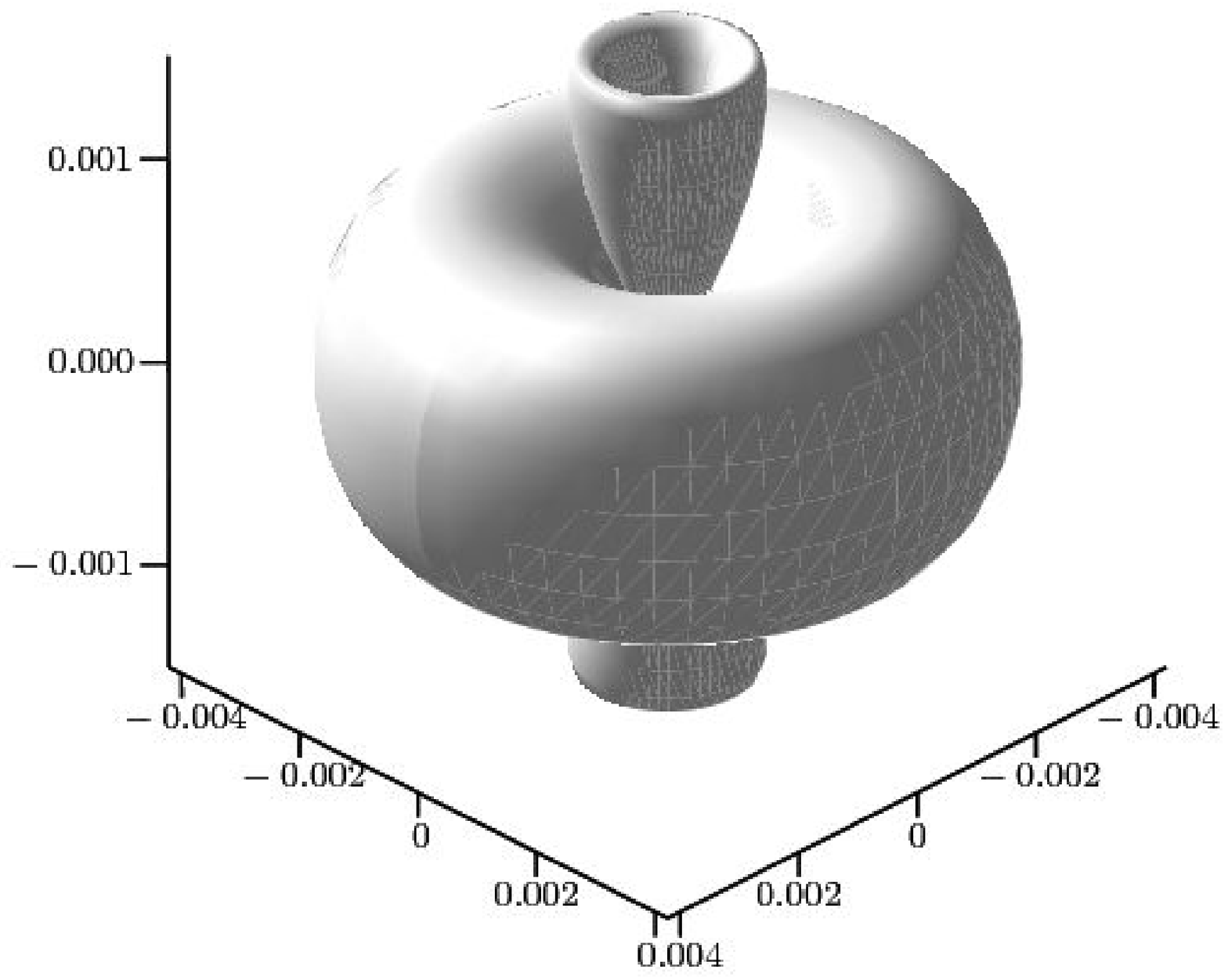}
\includegraphics[scale=0.27]{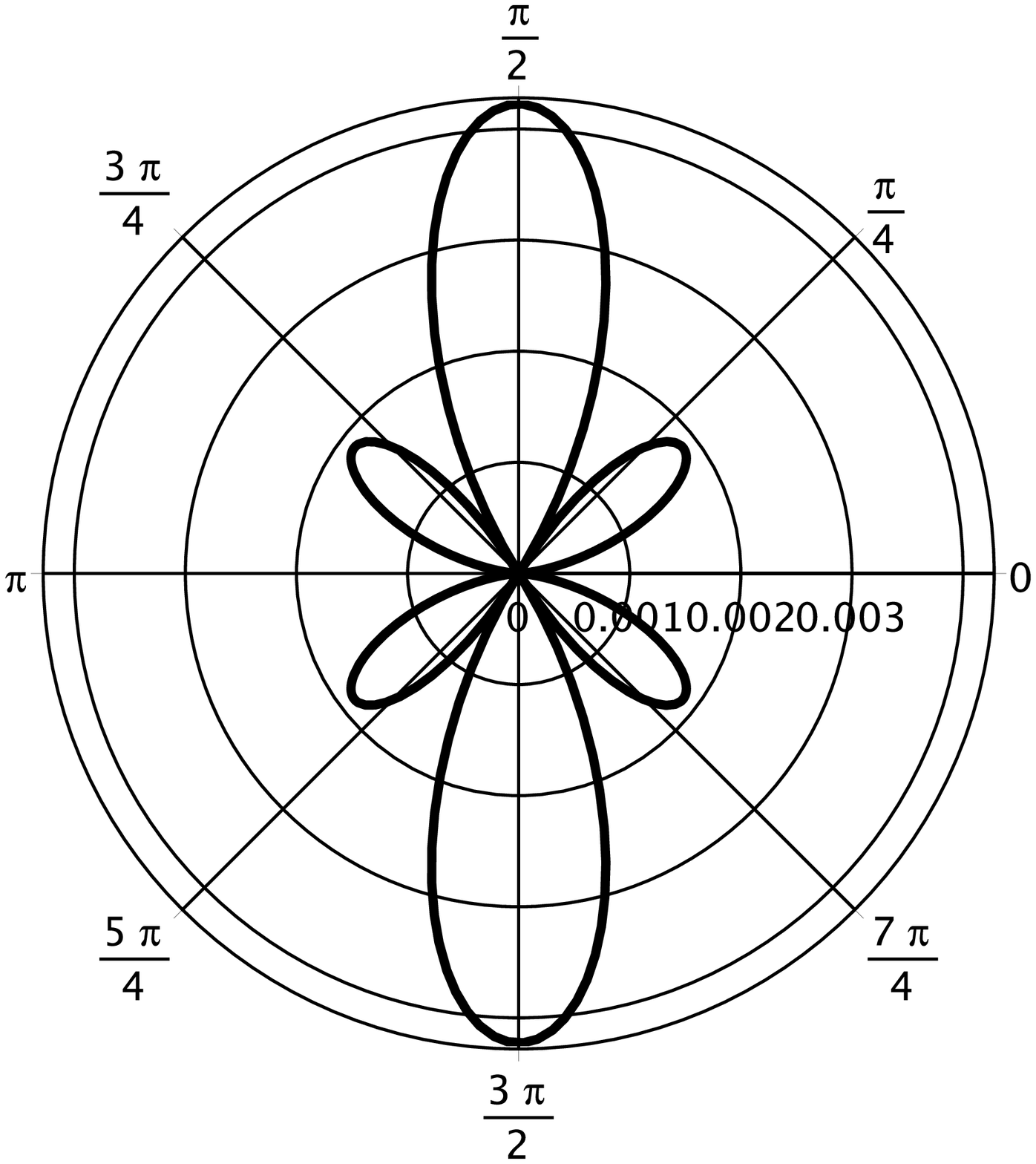}
\caption{From top to bottom we show the three and bi-dimensional polar plots (plane $\phi=0$) of $\Psi_4$ for the axisymmetric case ($c=0$). The lobe structure is symmetric as expected from the time symmetry condition imposed on the initial data. Also, it can be seen that the quadrupole mode is dominant.}
\end{figure}

We now turn to the problem of gravitational waveform extraction, whose accurate calculation is the most relevant problem in numerical relativity. The spin-weighted scalar $\Psi_4$ defined in the Newman-Penrose formalism \cite{NP} provides a measure of the outgoing gravitational radiation \cite{alcubierre_2,baumgarte}. Therefore, we shall determine the pattern of radiation perceived by a distant observer from the source by examining the dominant terms resulting the limit $r \rightarrow \infty$ of $\Psi_4$. This scalar is expressed by the following projection of the Weyl tensor $C_{\mu\nu\alpha\beta}$:

\begin{eqnarray}
\Psi_4 = C_{\mu\nu\alpha\beta} l^{\mu}\bar{m}^{\nu}l^\alpha \bar{m}^\beta,\label{eq30}
\end{eqnarray}

\noindent where $l^\mu$ and $\bar{m}^\nu$ belong to the null tetrad basis adopted by Bernstein et al \cite{bernstein} and shown in the Appendix. The complete expressions for the real and imaginary parts of $\Psi_4$ in the initial slice are also shown in the Appendix. Taking into account the asymptotic expression of the conformal factor, the choices of unit lapse and zero shift, and since the function $q(r,\theta,\phi)$ decays exponentially with $r$, we have found that $\Psi_4 \sim \mathcal{O}(r^{-3})$ at the initial slice, instead the typical decay $\mathcal{O}(r^{-1})$ which characterizes the wave zone. We attribute this behavior to the condition of time symmetry demanding that $K_{jk}=0$ at the initial slice. %In this case, the term $\partial_{[j} K_{l]k}$ is part of the Weyl tensor (cf. \cite{baumgarte}) decays as $\mathcal{O}(r^{-1})$ at the asymptotic region, while those terms that depend on the spatial metric and its derivatives decays faster.Therefore, in a dynamic setting we expect to recover the standard asymptotic behavior of $\Psi_4$ at subsequent slices.
We may recover the standard asymptotic behavior of $\Psi_4$ at subsequent slices in a dynamical setting. Notwithstanding this fact, we can consider the dominant terms of $\Psi_4$ for large $r$ as characterizing the structure of gravitational wave distribution at the initial slice. The asymptotic expression of $\Psi_4$ is,

\begin{eqnarray}
\lim_{r \rightarrow \infty} r^3 \Psi_4 &=& -(r \Psi)_{,\theta\theta} + (r \Psi)_{,\theta}\cot(\theta) + (r \Psi)_{,\phi\phi}\csc^2\theta +\nonumber  \\
&+& \frac{3 i}{2 \sin \theta}\left[(r \Psi)_{,\theta\phi} -(r \Psi)_{,\phi}\cot\theta \right].\label{eq31}
\end{eqnarray}

We have considered the approximated conformal factor with $N_x=20,N_y=16$ corresponding to the first initial data and obtained the expressions for the above asymptotic real and imaginary parts of $\Psi_4$. Hereafter, these pieces are denoted by $\Psi_4^{\mathrm{real}}$ and $\Psi_4^{\mathrm{im}}$, respectively. We also have set $n=4$, $\eta_0=1$ and $\sigma=1$ reducing the parameter space to $(A_0,c)$. In general, the angular distribution of $\Psi_4$ has the same symmetric of the conformal factor, but it depends on the parameters $(A_0,c)$. For the sake of convenience, all graphs correspond to $A_0=-1$, and the parameter $c$ can assume one of these values: $-2,-1,0,1$. In Fig. 7 we show the three and bi-dimensional polar plots of $\Psi_4$ for the axisymmetric case ($c=0$). In Fig. 8 we present a sequence of three and bi-dimensional polar plots of $\Psi_4^{\mathrm{real}}$. It can be seen  the role of the parameter $c$ in changing the angular or lobe structure of the pattern. On the other hand, the angular pattern of $\Psi_4^{\mathrm{im}}$, shown in Fig. 9, does not change significantly with respect to $c$.

\begin{figure*}
\includegraphics[height=6.0cm,width=6.0cm]{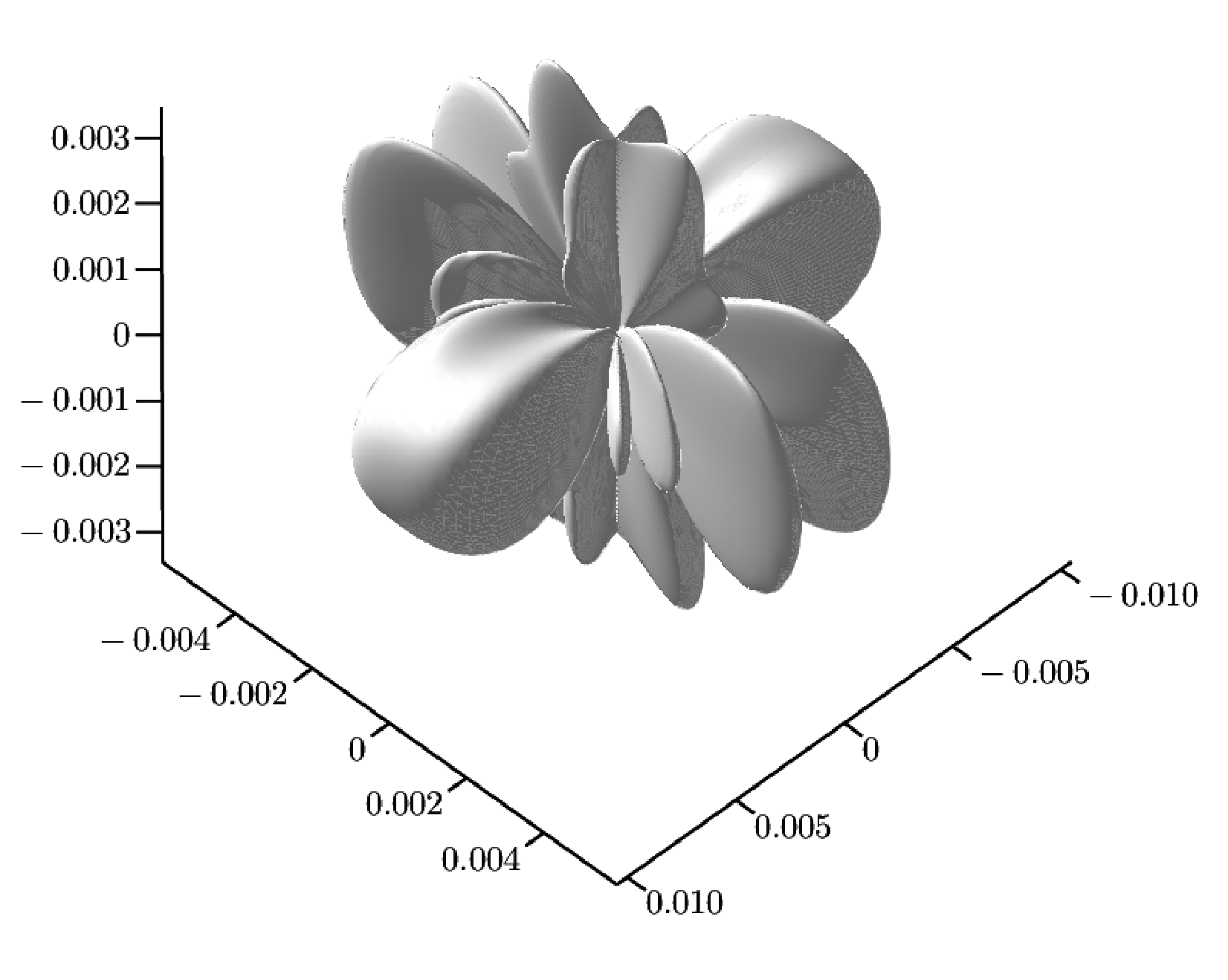}\includegraphics[height=6.0cm,width=6.0cm]{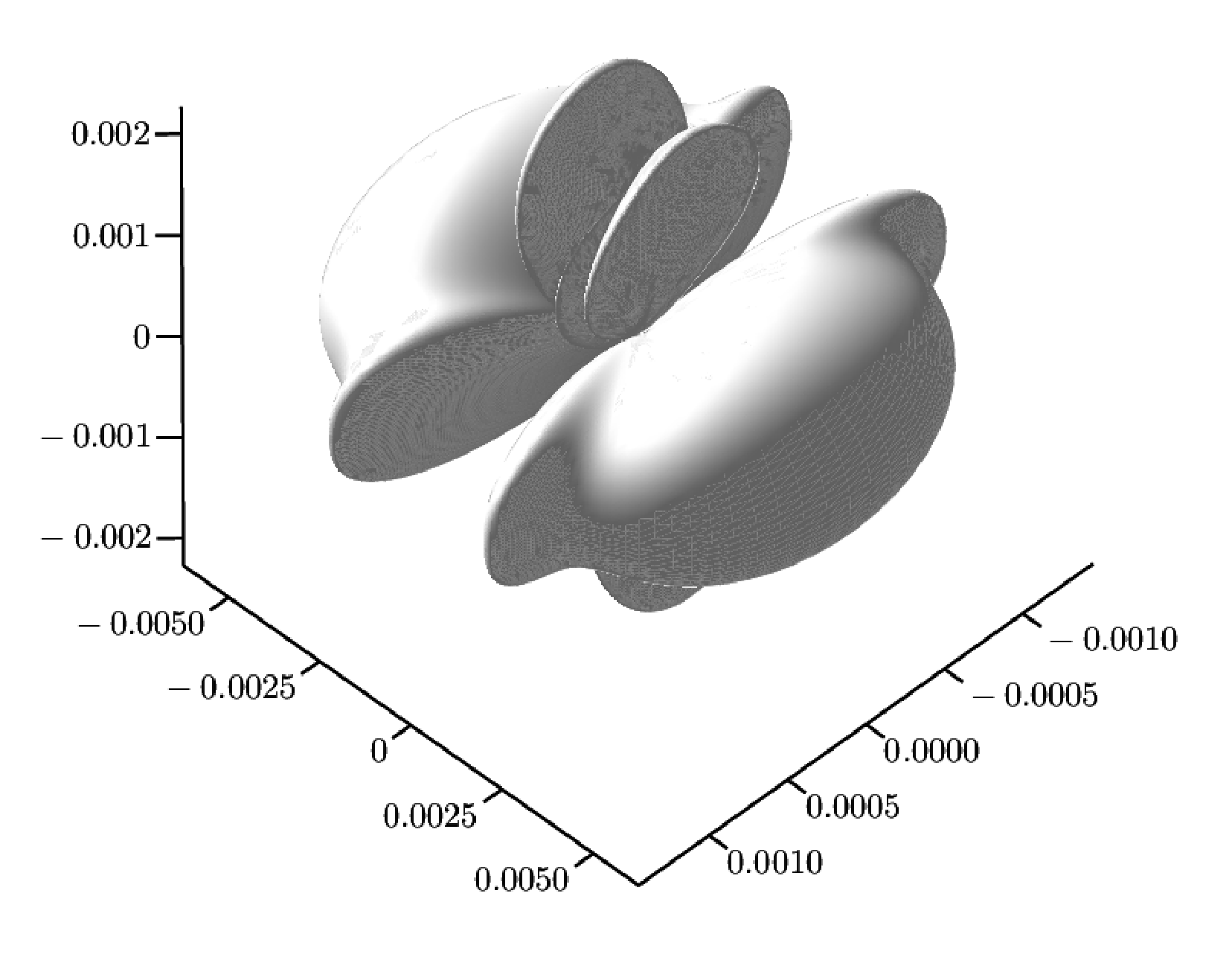}\includegraphics[height=5.5cm,width=6.0cm]{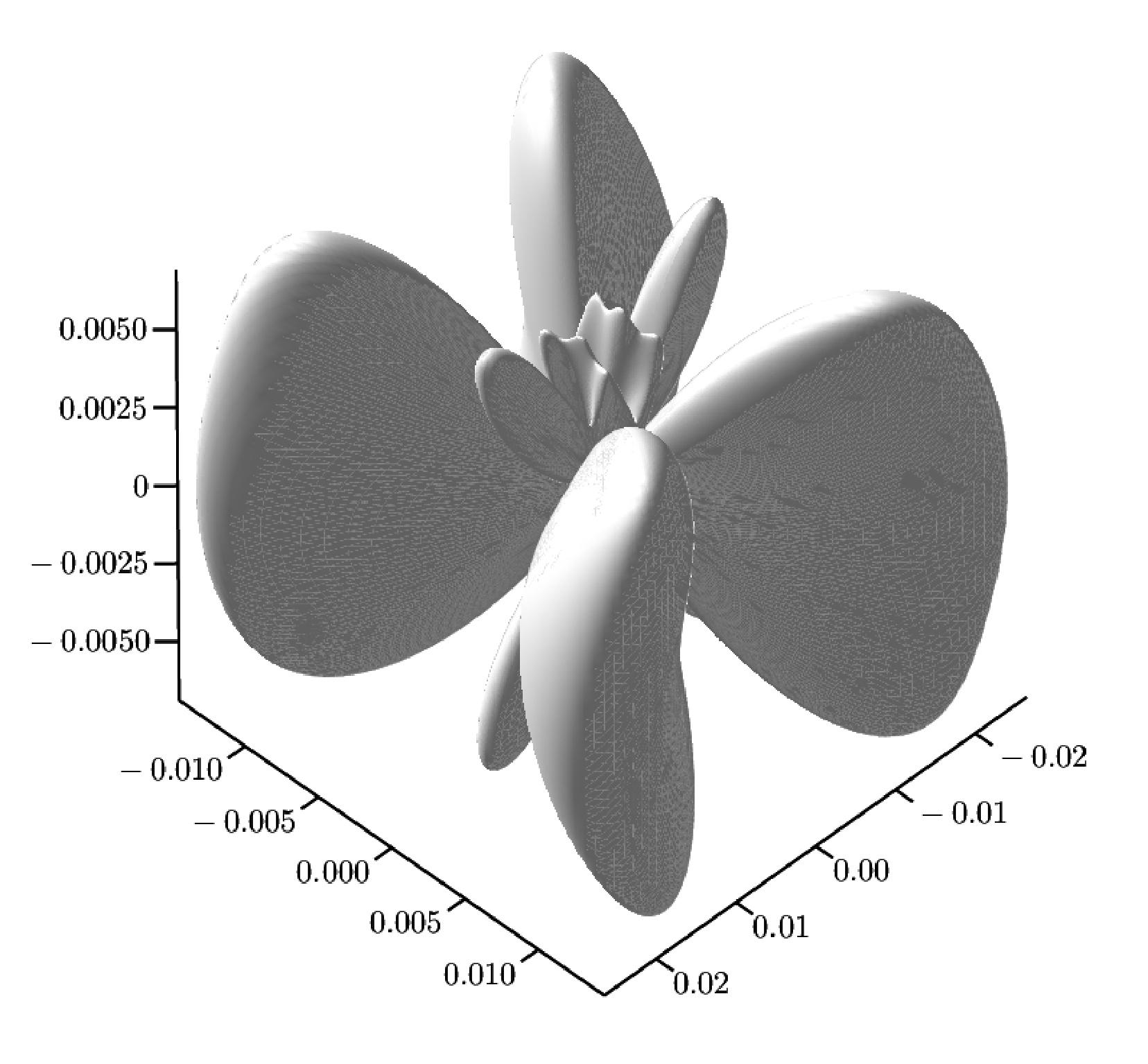}
\includegraphics[height=5.0cm,width=5.5cm]{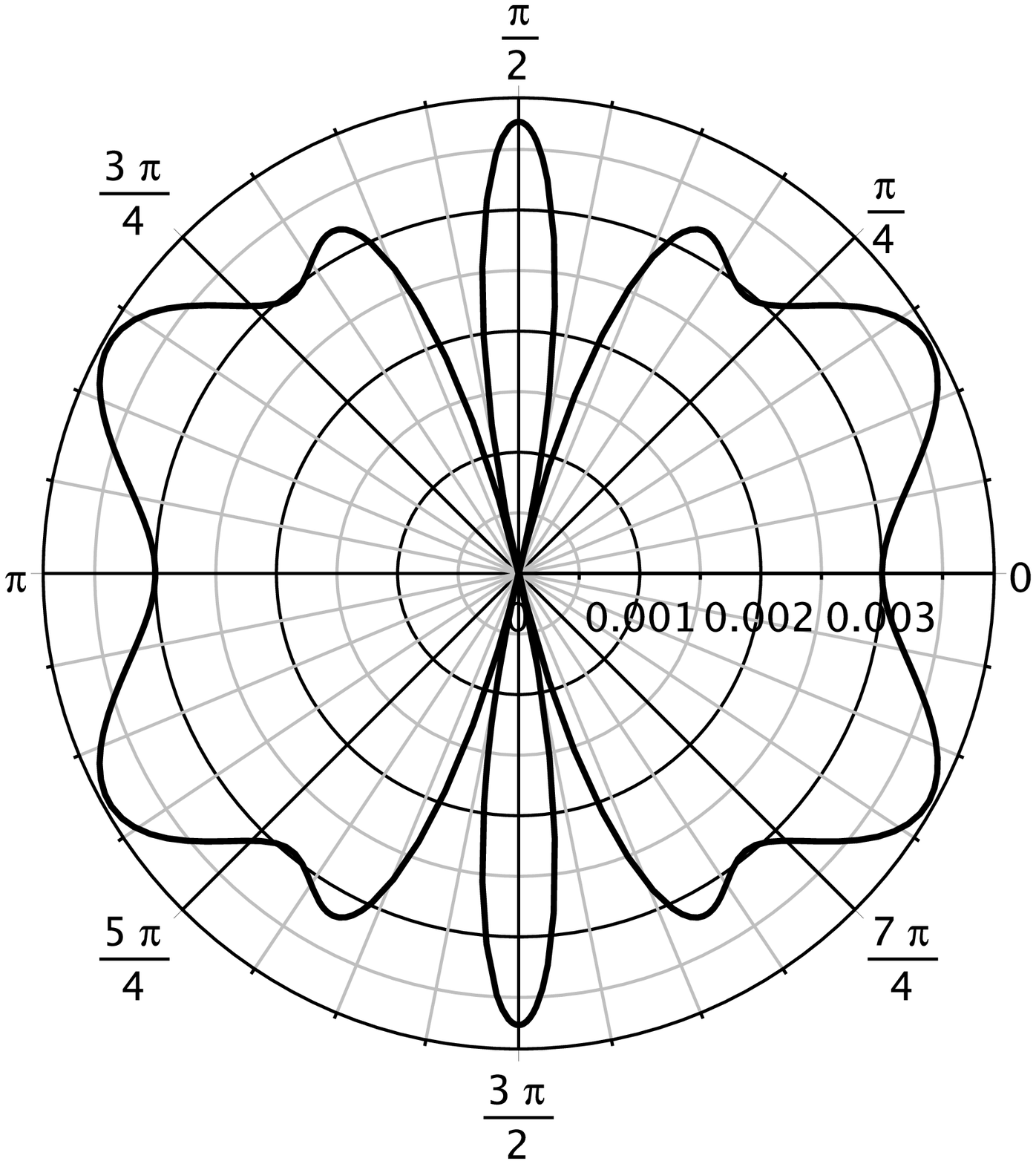}\includegraphics[height=5.0cm,width=5.5cm]{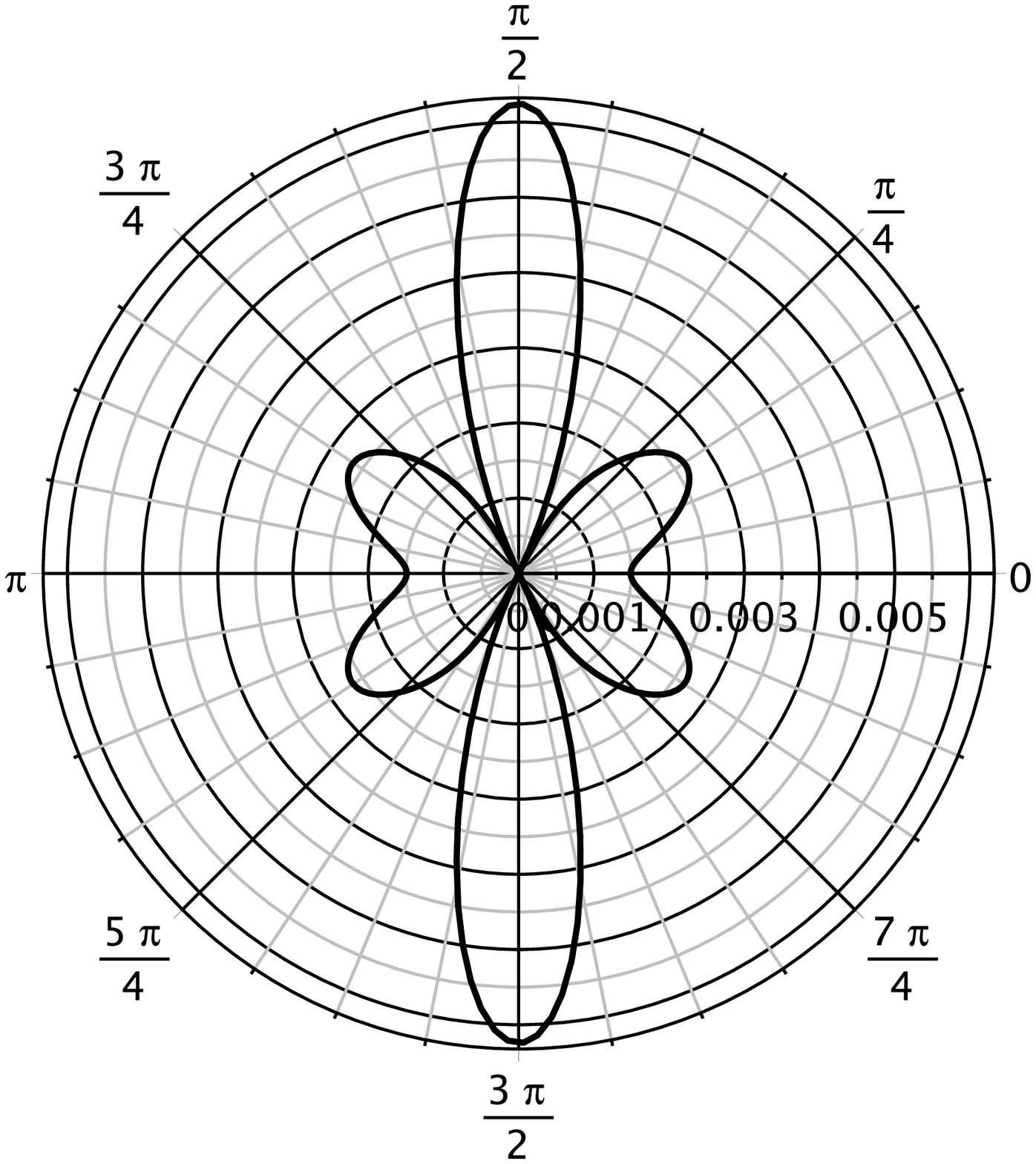}\includegraphics[height=5.0cm,width=5.5cm]{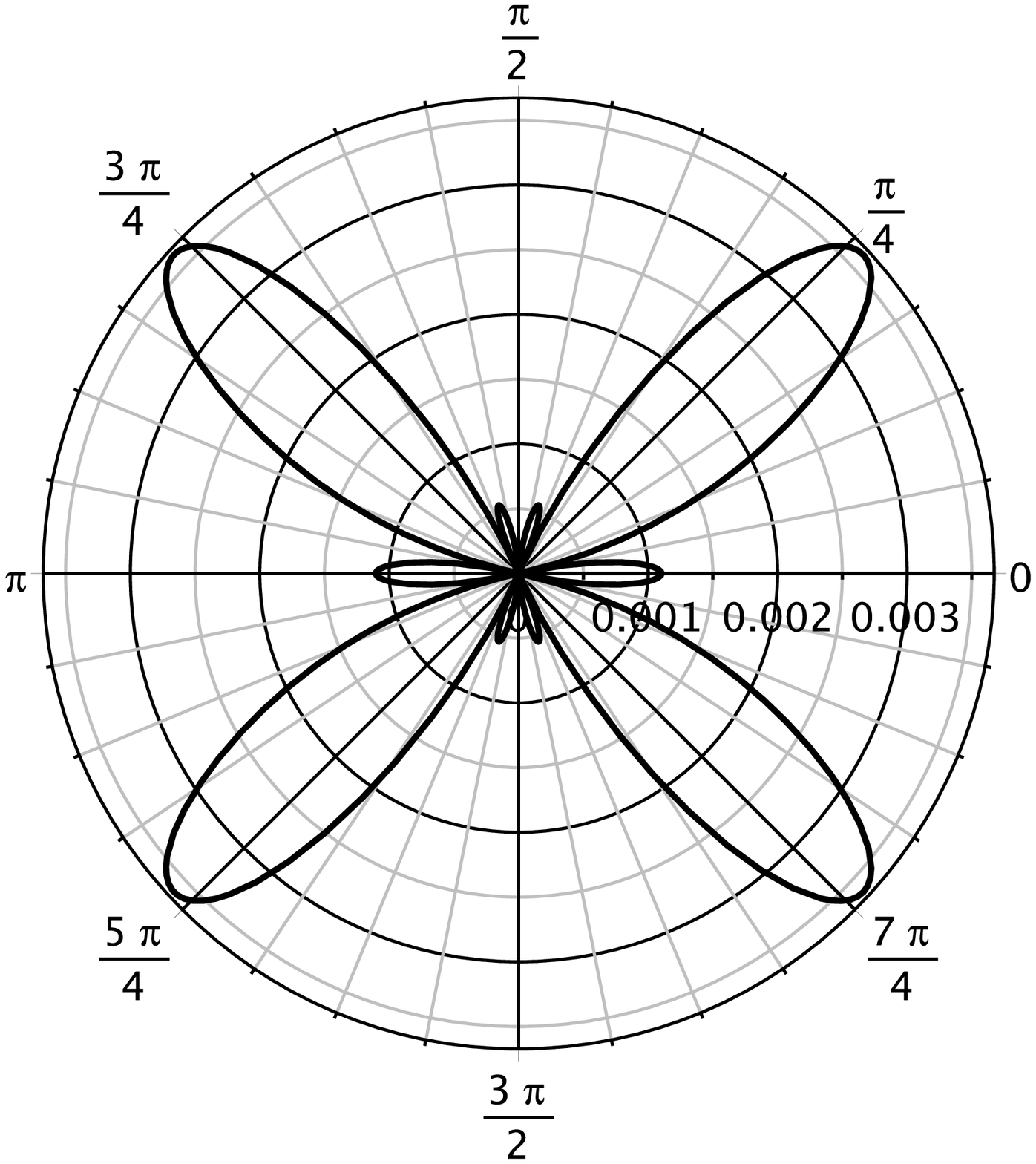}
\caption{Sequence of three and bi-dimensional polar plots (plane $\phi=\pi/2$) $\Psi_4^{\mathrm{real}}$ with $c=-2,-1,1$, from left to right. In all cases $A_0=-1,n=4,\eta_0=1,\sigma=1$ in the first initial data. The deviation from axisymmetry produces a rich multipole structure.}
\end{figure*}

\begin{figure}[htb]
\includegraphics[scale=0.25]{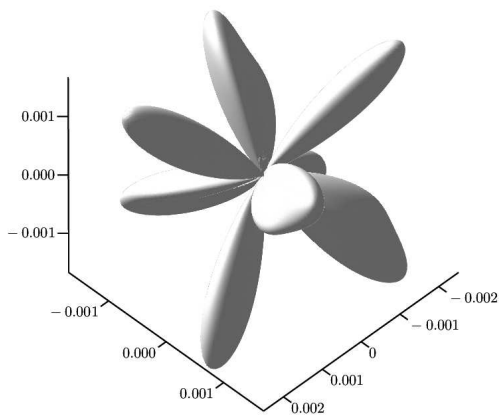}
\caption{The structure of $\Psi_4^{\mathrm{im}}$ does not change considerably with $c,A_0$.}
\end{figure}

%---------------------------------------------------------------------------------------------------------------------------------------------------------------------------------------------------------------------------------------
\section{Final remarks}%
%---------------------------------------------------------------------------------------------------------------------------------------------------------------------------------------------------------------------------------------

In this work, we have implemented a Galerkin-Collocation spectral algorithm to solve the Hamiltonian constraint corresponding to three-dimensional distorted black holes. These configurations can describe two plausible astrophysical situations: the late stages of black hole coalescence or the interaction of a black hole with a cloud of gravitational waves.

We have solved the Hamiltonian constraint in the realm of the inversion method as a direct generalization of the axisymmetric case \cite{deol_rod_idata}, and also implemented the Galerkin-Collocation version of the puncture method. According to this method, the black hole interior is not excised, and the spatial domain $0 < r < \infty$ is taken into consideration. We have developed a simple version of the domain decomposition when compared with other versions found in the literature \cite{pfeiffer_CPC,ansorg_1,pfeiffer_thesis,pfeiffer_bh_gw,brown_lowe,grandclement,ansorg_bbh,ansorg_07}. In this case, we have divided the spatial domain into two regions, $0 < r \leq r_0$ and $r \geq r_0$. The numerical experiments indicated that the boundary $r=r_0$ is better placed at the black hole throat. For the sake of comparison, we have exhibited the convergence of the ADM mass for both, puncture with domain decomposition and inversion methods, in Fig. 3.

In both methods the conformal factor is expressed in as a series expansion of radial and angular basis functions constituted by, respectively, a suitable combination of the rational Chebyshev polynomials and spherical harmonics.  With respect to the angular basis, we, metaphorically speaking, are dealing with a double-edged sword. These functions are the best basis on the sphere offering exponential convergence for regular functions defined on the sphere. On the other hand, spherical harmonics are more complicated than any other basis functions because they are two-dimensional \cite{boyd}.

It is worth commenting the similarities and differences of the present numerical implementation with the one by Pfeiffer et al. \cite{pfeiffer_CPC} since it is a spectral code with similar basis functions. The first difference is that we have adopted radial basis functions that are suitable linear combinations of the pure Chebyshev polynomials, instead of pure Chebyshev polynomials of Ref. \cite{pfeiffer_CPC}. The linear combination is such that each basis function satisfies de boundary conditions. As pointed out by Heinrichs \cite{heinrichs}, a combination of pure Chebyshev polynomials produces accurate results due to lower accumulated round-off error when solving higher order differential equations. Another difference comes from the use of mappings. We have considered the algebraic map \cite{boyd} that is more adequate to describe functions with algebraic asymptotic behavior, which is the case of the conformal factor. In spite of considering the same angular basis of Ref. \cite{pfeiffer_CPC}, we have employed the Galerkin method with numerical integration, GN-I, in the angular domain instead of the Collocation method. However, the Collocation method was used in the radial domain. Finally, we point out that in Ref. \cite{pfeiffer_CPC} the spatial domain is divided into several regions, whereas we have considered only two regions. The simplicity is due to the specific initial data problem we are dealing.

We remark that the conformal factor is defined in the entire spatial domain under consideration in each method. This feature provides a natural and simple determination of the ADM mass by calculating the asymptotic spatial limit of the term $r^2 \partial \Psi/\partial r$ (cf. Eq. 28). In addition, we have examined the angular pattern associated to the dominant term of the spin-weighted scalar $\Psi_4$ at the spatial infinity. In the present case we have found that $\Psi_4 \simeq \mathcal{O}(r^{-3})$. We can interpret such patterns as the indicators of the gravitational waves at a large distance from the distorted black hole. The angular patterns present a rich structure that depend upon the parameters of the initial data, and can be understood as gravitational-wave fingerprints distorted black holes.

The Galerkin-Collocation method is a viable alternative to solve the initial data problem of distorted three-dimensional black holes. The next natural direction of the present research is to study the dynamics of distorted black holes. There are two issues we will be focusing, namely, the gravitational wave templates produced in this dynamical process and the efficiency of the gravitational wave extraction. The previous works on the dynamics of distorted non-rotating \cite{abrahams,camarda} and rotating black holes \cite{new_pfeifer} have not discussed in details these issues.

\section*{Acknowledgements}

The authors acknowledge the financial support of the Brazilian agencies CNPq, CAPES and FAPERJ.

%\vspace{0.7cm}

\appendix*

\section{}

We present here null tetrad basis assuming unit lapse and zero shift on the initial slice, and the with the three metric given by Eq. (\ref{eq5}). Then,

%{\small
\begin{eqnarray}
l^\mu &=& \frac{1}{\sqrt{2}}\left(1,\frac{-1}{\Psi^2 \mathrm{e}^q},0,0\right),\\
k^\mu &=& \frac{1}{\sqrt{2}}\left(1,\frac{1}{\Psi^2 \mathrm{e}^q},0,0\right), \\
m^\mu &=& \frac{1}{\sqrt{2}}\left(0,0,\frac{1}{r\Psi^2 \mathrm{e}^q},\frac{i}{r\Psi^2 \sin \theta}\right),\\
\bar{m}^\mu &=& \frac{1}{\sqrt{2}}\left(0,0,\frac{1}{r\Psi^2 \mathrm{e}^q},\frac{-i}{r\Psi^2 \sin \theta}\right).
\end{eqnarray}
%}

\noindent The real and imaginary parts of the scalar $\Psi_4$ are,

\begin{widetext}
\begin{eqnarray}
\Psi_4^{\mathrm{(real)}} &=& \frac{1}{\Psi^4 r^2 \mathrm{e}^{2q}}\left(\frac{1}{2}q_{,\theta} \cot \theta - \frac{\Psi_{,r}}{\Psi} r^2 q_{,r} + \frac{\Psi_{,\theta}}{\Psi} q_{,\theta} - r q_{,r} - \frac{\Psi_{,\theta\theta}}{\Psi} - \frac{1}{2} q_{,\theta\theta} + 3\frac{\Psi_{,\theta}^2}{\Psi^2} + \frac{\Psi_{,\theta}}{\Psi} \cot \theta - \frac{1}{2} r^2 q_{,rr}\right) - \nonumber \\
& & \frac{1}{\Psi^4 r^2 \mathrm{e}^{2q} \sin^2\theta}\left(3\frac{\Psi_{,\phi}^2}{\Psi^2} - \frac{1}{2} q_{,\phi\phi} + \frac{\Psi_{,\phi}}{\Psi}q_{,\phi} -\frac{\Psi_{,\phi\phi}}{\Psi}\right)
\nonumber \\
\Psi_4^{\mathrm{(im)}} &=& -\frac{1}{\Psi^4 r^2 \mathrm{e}^{q} \sin \theta}\left(-\frac{3\Psi_{,\theta\phi}}{2\Psi} - \frac{3}{4} q_{,\theta\phi} + \frac{3\Psi_{,\theta}}{2\Psi} q_{,\phi} + \frac{3\Psi_{,\phi}}{2\Psi} \cot \theta + \frac{3}{4} q_{,\phi} \cot \theta + \frac{9}{2} \frac{\Psi_{,\theta}\Psi_{,\phi}}{\Psi^2}\right).
\end{eqnarray}
\end{widetext}


\begin{thebibliography}{99}

\bibitem{pfeiffer_3d} Harald P. Pfeiffer, Class. Quant. Grav., \textbf{29}, 124004 (2012).

\bibitem{bernstein} D. Bernstein, D. Hobill, E. Seidel and L. Smarr, \textit{Phys. Rev. D} \textbf{50}, 3760 (1994).

\bibitem{brandt} S. Brandt, K. Camarda, E. Seidel and R. Takahashi, Class. Quant. Grav., \textbf{20}, 1 (2003).

\bibitem{deol_rod_idata} H. P. de Oliveira and E. L. Rodrigues, \textit{Phys. Rev. D} \textbf{86}, 064007 (2012).

\bibitem{kidder} L. E. Kidder and L. S. Finn, \textit{Phys. Rev. D} \textbf{62}, 084026 (2000).

\bibitem{id_finite_dif} M. Choptuik and W. G. Unruh, \textit{Gen. Relativ. Gravit.} \textbf{18}, 813 (1986); G. B. Cook, Ph.D. thesis, University of North Carolina, Chapel Hill (1990); D. Bernstein, D. Hobill, E. Seidel, and L. Smarr, \textit{Phys. Rev.} \textbf{D} 50, 3760 (1994).

\bibitem{pfeiffer_CPC} Harald P. Pfeiffer, Lawrence E. Kidder, Mark A. Scheel and Saul Teukolsky, Comp. Phys. Commun. \textbf{152}, 253 (2003).

\bibitem{ansorg_1} Marcus Ansorg, Bernd Brugmann and Wolfang Tichy, Phys. Rev. D \textbf{70}, 064011 (2004).

\bibitem{pfeiffer_thesis} Harald P. Pfeiffer, \textit{Initial data for black hole evolutions}, PhD thesis, preprint arXiv: gr-qc/0510016.

\bibitem{pfeiffer_bh_gw} Harald P. Pfeiffer, Lawrence E. Kidder, Mark A. Scheel and Deirdre Shoemaker, Phys. Rev. D \textbf{71}, 024020 (2005).

\bibitem{schinkel} David Schinkel, Marcus Ansorg and Rodrigo Panoso Macedo, \textit{Initial data for perturbed kerr black holes on hyperboloidal slices}, prepring arXiv: gr-qc/1301.6984.

\bibitem{brandt_brugmann} S. Brandt and B. Brugmann, Phys. Rev. Lett. \textbf{78}, 3606 (1997).

\bibitem{brown_lowe} J. David Brown and Lisa L. Lowe, Phys. Rev. D \textbf{70}, 124014 (2004).

\bibitem{grandclement} Phillipe Grandclement, Eric Gourgoulhon and Silvano Bonazzola, Phys. Rev. D \textbf{65}, 044021 (2002).

\bibitem{ansorg_bbh} Marcus Ansorg, Phys. Rev. D \textbf{72}, 024018 (2005).

\bibitem{ansorg_07} Marcus Ansorg, Class. Quant. Grav. \textbf{24}, S1-14 (2007).

\bibitem{pfeiffer2} Harald P Pfeiffer, Duncan A Brown, Lawrence E Kidder, Lee Lindblom, Geoffrey Lovelace and Mark A Scheel, Class. Quantum Grav. 24 S59 (2007).

\bibitem{foucart} Francois Foucart, Lawrence E. Kidder, Harald P. Pfeiffer, and Saul A. Teukolsky, Phys. Rev. D \textbf{77}, 124051 (2008).

\bibitem{ruchlin} Ian Ruchlin, James Healy, Carlos O. Lousto, Yosef Zlochower, arXiv:1410.8607 [gr-qc] (2014).

\bibitem{lovelace}  Geoffrey Lovelace, Robert Owen, Harald P. Pfeiffer, and Tony Chu, Phys. Rev. D \textbf{78}, 084017 (2008).

\bibitem{koutarou}  Koutarou Kyutoku, Masaru Shibata, and Keisuke Taniguchi, Phys. Rev. D \textbf{90}, 064006 (2014).

\bibitem{deol_rod_bondi} H. P. de Oliveira and E. L. Rodrigues, Class. Quant. Grav, \textbf{28}, 235011 (2011).

\bibitem{deol_rod_RT} H. P. de Oliveira, E. L. Rodrigues and J. E. F. Skea, \textit{Phys. Rev.} \textbf{D} 84, 044007 (2011).

\bibitem{adm} R. Arnowitt, S. Deser and C. W. Misner, {The dynamics of general relativity}, in \textit{Gravitation: An Introduction to Current Research}, ed. L. Witten (Wiley, New York, 1962), p.227.

\bibitem{york} J. W. York Jr., {The initial value problem and dynamics}, in {\textit Gravitational Radiation}, eds. N. Deruelle and T. Piran (North-Holland, Amsterdan, 1983).

\bibitem{brill} D. Brill, \textit{Ann. Phys. (N.Y.)} \textbf{7}, 466 (1959).

\bibitem{boyd} J. P. Boyd, \textit{Chebyshev and Fourier Spectral Methods} (Dover Publications, New York, 2001).

\bibitem{canuto} C. Canuto, M. Y. Hussaini, A. Quarteroni and T. A. Zang, \textit{Spectral Methods in Fluid Dynamics} (Springer-Verlag, Berlin, Germany; Heidelberg, Germant, 1988); \textit{Spectral Methods: Fundamentals in Single Domains} (Springer-Verlag, Berlin, Germany; Heidelberg, Germant, 2006); Roger Peyert, \textit{Spectral Methods for Incompressible Viscous Flow}, Springer(2001).

\bibitem{fornberg} B. Fornberg, \textit{A Pratical Guide to Pseudospectral Methods}, Cambridge Monographs on Applied and Computational Mathematics, Cambrige University Press (1998).

\bibitem{peyret} Roger Peyret, \textit{Spectral Methods for Incompressible Viscous Flow}, Applied Mathematical Sciencies, 148, Springer-Verlag (2000).

\bibitem{finlayson} B. A. Finlayson, \textit{The Method of Weighted Residuals and Variational Principles} (Academic Press, New York, 1972).

\bibitem{ADM_mass} N. \'O Murchadha and J. York, Phys. Rev. D\textbf{10}, 24345 (1974).

\bibitem{NP} E. T. Newman and R. Penrose, \textit{J. Math. Phys.} \textbf{3}, 566 (1962); \textit{J. Math. Phys.} \textbf{4}, 998 (1963).

\bibitem{alcubierre_2} M. Alcubierre, \textit{Introduction to 3+1 Numerical Relativity}, Oxford University Press (2008).

\bibitem{baumgarte} T. Baumgarte and S. L. Shapiro, \textit{Numerical Relativity - Solving the Einstein's Equations on the Computer}, Cambridge University Press (2010).

\bibitem{heinrichs} W. Heinrichs, J. Comp. Phys. \textbf{59}, 103 (1989); J. Scient. Comp. \textbf{6}, 1 (1991); SIAM J. Scient. and Stat. Comp. \textbf{12}, 1162 (1991).

\bibitem{abrahams} A. Abrahams, D. Bernstein, D. Hobill, E. Seidel, and L. Smarr, Phys. Rev. D \textbf{45}, 3544 (1992). A. M. Abrahams and C. R. Evans, Phys. Rev. D \textbf{42}, 2585 (1990). D. Bernstein, D. Hobill, E. Seidel, L. Smarr, and J. Towns, Phys. Rev. D \textbf{50}, 5000 (1994).

\bibitem{camarda} K. Camarda and E. Seidel, Phys. Rev. D \textbf{57}, R3204 (1998). J. Baker, S. Brandt, M. Campanelli, C. O. Lousto, E. Seidel and R. Takahashi, Phys. Rev. D \textbf{62}, 127701 (2000).

\bibitem{new_pfeifer} Tony Chu, Harald P. Pfeiffer and Michael I. Cohen, Phys. Rev. D \textbf{83}, 104018 (2011), gr-qc 1011.2601.

\end{thebibliography}
\end{document}